# The Bright Side and the Dark Side of Hybrid Organic Inorganic Perovskites

*Wladek Walukiewicz[1,2],\*, Shu Wang[3], Xinchun Wu[4], Rundong Li[3], Matthew P. Sherburne[5], Bo Wu[6], Tze Chien Sum[7], Joel W. Ager[1,2,5], and Mark D. Asta[2,5]*

[1]Berkeley Educational Alliance for Research in Singapore (BEARS), Ltd., 1 Create Way, 138602, Singapore

[2]Materials Sciences Division, Lawrence Berkeley National Laboratory, Berkeley, CA, 94720, USA

[3]College of Materials Science and Opto-Electronic Technology, University of Chinese Academy of Sciences, Beijing, 100049, P. R. China

[4]Department of Chemistry Science, University of Chinese Academy of Sciences, Beijing, 100049, P. R. China

[5]Department of Materials Science and Engineering, University of California, Berkeley, California 94720, USA

[6]Guangdong Provincial Key Laboratory of Optical Information Materials and Technology & Institute of Electronic Paper Displays, South China Academy of Advanced Optoelectronics, South China Normal University, Guangzhou 510006, P. R. China

[7]Division of Physics and Applied Physics, School of Physical and Mathematical Sciences, Nanyang Technological University, 21 Nanyang Link, Singapore 637371

\*Phone: +1 510 486 5329, Email: w_walukiewicz@lbl.gov

Abstract

The previously developed bistable amphoteric native defect (BAND) model is used for a comprehensive explanation of the unique photophysical properties and for understanding the remarkable performance of perovskites as photovoltaic materials. 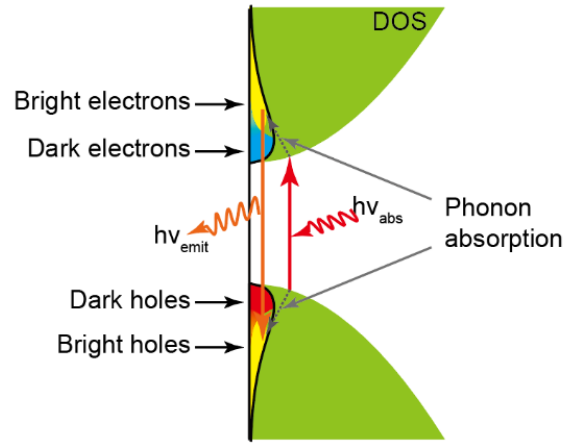 It is shown that the amphoteric defects in donor (acceptor) configuration capture a fraction of photoexcited electrons (holes) dividing them into two groups: higher energy bright and lower energy dark electrons (holes). The spatial separation of the dark electrons and the dark holes and the k-space separation of the bright and the dark charge carriers reduce electron hole recombination rates, emulating the properties of an ideal photovoltaic material with a balanced, spatially separated transport of electrons and holes. The BAND model also offers a straightforward explanation for the exceptional insensitivity of the photovoltaic performance of polycrystalline perovskite films to structural and optical inhomogeneities. The blue-shifted radiative recombination of bright electrons and holes results in a large anti-Stokes effect that provides a quantitative explanation for the spectral dependence of the laser cooling effect measured in perovskite platelets.

1. Introduction

Recent years have witnessed unprecedented progress in applications of hybrid organic-inorganic perovskites (HOIPs) for solar power conversion. Intense efforts of numerous research groups have led to spectacular improvements in the perovskite based solar cell power conversion efficiency from initial 3.8 %[1] to the current record of more than 25 %[2]. Since this progress was achieved using simple, inexpensive materials and synthesis methods it has generated significant interest in commercial applications of this thin film PV technology[3]. Also, it has stimulated an enormous research effort aimed at understanding the basic properties of the perovskites. Various aspects of the perovskite related research have been extensively discussed in several reviews[4-10].

It was recognized quite early that the HOIPs cannot be viewed as standard semiconductors as they show a range of unusual properties not found in other semiconductor materials. Thus, most surprisingly, cladding of an insulating perovskite film with electron and hole accepting layers produces a p-i-n PV with the open circuit voltage determined by the band gap of the perovskite material rather than the Fermi level difference between the electron and hole transporting layers[6, 11].

In our first paper[12] we have proposed that the exceptional properties of HOIPs as PV absorbers can be explained by bistable amphoteric native defects (BANDs), a class of defects that can undergo a transformation between shallow donor and shallow acceptor configuration. The BANDs model accounts for a variety of the phenomena observed in HOIPs including formation of p-i-n junction PVs as well as hysteresis of J-V characteristics and reversible UV degradation of PV devices[12]. There is growing evidence showing that the iodine site defects play an important role in determining the photophysical properties and stability of MAPbI$_3$[13, 14]. In the amphoteric defect model[12] an iodine vacancy donor, $D^+ \equiv V_I^+$ with MA dangling bonds undergoes a structural relaxation in which the neighboring MA molecule moves into the vacant I site forming MA vacancy like acceptor complex $A^- \equiv [V_{MA} + MA_I]^-$ with iodine dangling bonds. This fully reversible defect transformation depends on the quasi-Fermi energy as it requires participation of free carriers[12].

There are other properties of HOIPs which are not satisfactorily understood. Extensive experimental studies have demonstrated that the HOIPs exhibit extraordinary optical and charge transport properties that cannot be easily explained in terms of standard semiconductor models. For example, unexpectedly long and balanced diffusion lengths[15, 16] for photoexcited electrons and holes have been observed and an exceptionally good PV performance has been realized in polycrystalline perovskites which have very large spatial and temporal variations of photoluminescence[17]. The findings that a poorer crystallinity and a larger concentration of structural defects do not affect perovskite material PV performance was further substantiated by experiments showing that the grain boundaries can actually play a beneficial role in collection of photoexcited charge carriers[18]. Also, detailed measurements of the optical properties of MAPbI$_3$ films[19] have shown a remarkably large blue shift of the PL energy relative to the absorption edge. This anti-Stokes shift was similar to the energy difference found for photons emitted from the surface and the bulk of single crystals[20, 21] and appeared to be consistent with experiments demonstrating optical cooling of perovskite platelets[22].

Separately, it was observed that low energy photoexcited electrons and holes appear to be protected from rapid recombination and contribute to a long lived photoconductivity[23] and to surprisingly long lifetimes for the photoluminescence and photoconductivity transients[24]. These findings suggest that HOIPs do not follow the standard radiative detailed balance rule that relates photon absorption to photon emission rates and imply that there is a disconnect between optical absorption and optical emission, with the bimolecular recombination rates dependent on extrinsic material parameters such as sample preparation conditions and/or sample thickness[25]. It has been argued that some of these properties of perovskites can be understood considering different combinations of four mechanisms[26]: (1) trapping and de-trapping by shallow defects, (2) transfer of photoexcited electrons from direct to indirect minimum, (3) polaron formation, and (4) photon recycling. A detailed evaluation of the different mechanisms tends to favor photon recycling[26] which has been used to analyze optical transients[27] as well as long distance carrier diffusion[28]. However, this mechanism requires an extremely efficient, close to 90% confinement, of randomly emitted photons in submicron thick polycrystalline films[27]. This requirement for very efficient photon confinement cannot be easily reconciled with the direct measurements of photon recycling showing a photon retention fraction of less than 0.5% in single HOIP crystals[29].

In this paper we present an entirely new, comprehensive explanation of the unique photophysical properties of HOIPs. We show that the Coulomb potentials of positively (negatively) charged BANDs in donor (acceptor) configuration separate the photoexcited charge carriers into bright electrons (holes) which can recombine radiatively and dark electrons (holes) which are captured by coulomb fields of the BANDs and are thus protected from recombination. The model is used to calculate photophysical properties of HOIPs under transient and steady state illuminations. It is shown that the storage of the photoexcitation energy by the dark carriers explains long lifetimes and long diffusion lengths and is the key reason for the outstanding performance of HOIP- based PV devices. Additionally, the model accounts for the large anti-Stokes shift of the photoluminescence and the laser cooling effect.

2. Separation of mobile charge carriers into bright and dark fractions

The essential feature of the bistable amphoteric native defects (BANDs) is that they can undergo a structural transformation between the donor and acceptor configuration. The dynamic balance between those two defect configurations is controlled by the concentrations of photoexcited electrons and holes[12]. These properties of BANDs were used to explain formation and the properties of p-i-n junction in perovskite films in contact with electron and hole transporting layers. Here we develop a model describing how the interactions between BANDs and mobile electrons and holes affect the optoelectronic properties of hybrid organic-inorganic perovskite materials.

In our calculations we consider MAPbI3 as a prototypical perovskite material with BANDs concentration of $2N_d$. Under equilibrium conditions the material is compensated and has high resistivity with the defects evenly distributed between donor and acceptor configurations, $N_D^+ = N_A^- = N_d$ [12]. Illumination of the sample with above the band gap energy monochromatic light produces electron-hole pairs with a narrow energy and k-wavector distribution, schematically shown in Figure 1(a). Electrons and holes assume an equilibrium thermal distribution shown in Figure 1(b) through absorption and emission of phonons. This thermalization is a fast process occurring on a picosecond time scale.

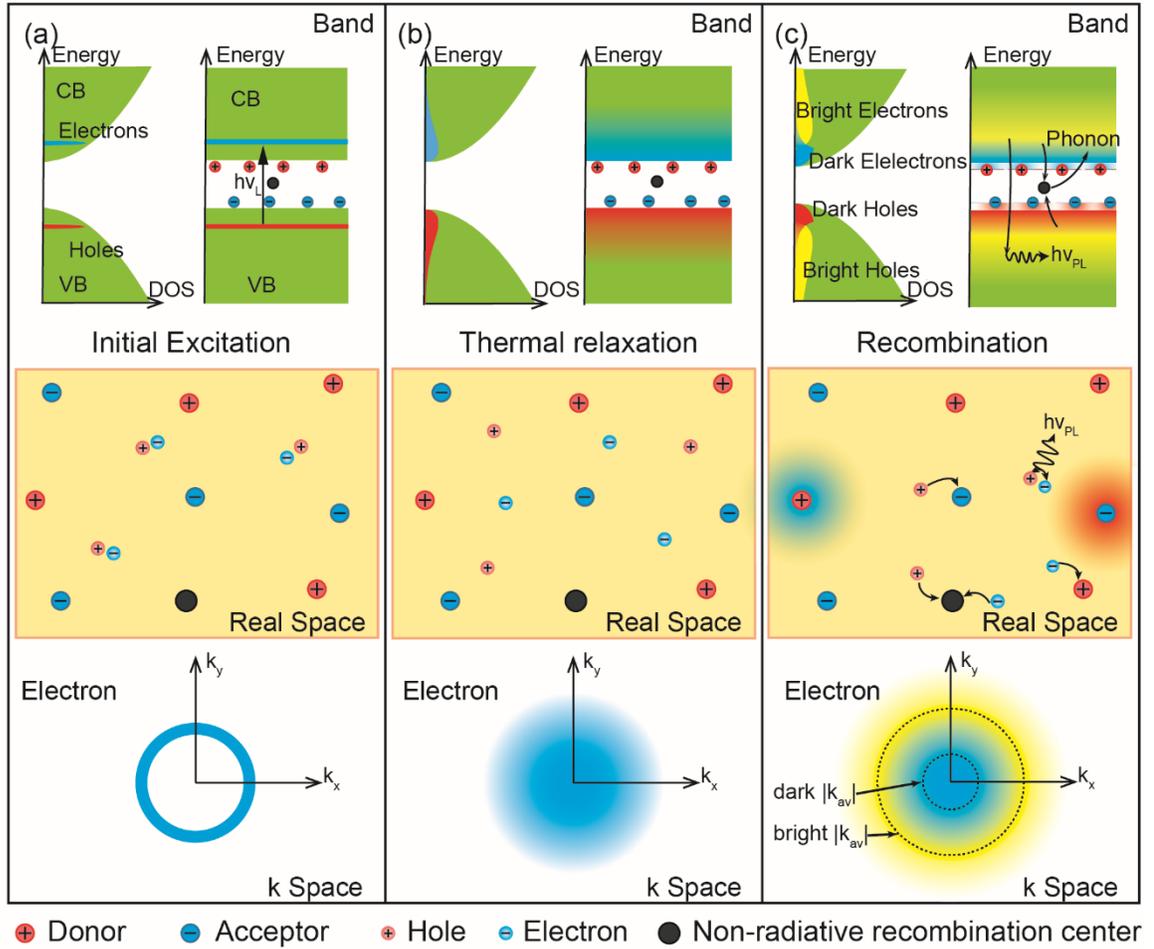

Figure 1. Separation of electrons and holes into bright and dark fractions illustrated in the density of states (upper left panels), energy (upper right panels), real space (middle panels) and k space (lower panels). (a) electrons and holes excited with monochromatic light into a narrow energy and k-wavevector range in the conduction and the valence band remain paired. (b) the electrons separate from holes and the charge carriers thermalize with the lattice by emitting and absorbing phonons. (c) The distributions of photoexcited electrons and holes calculated for the electron (hole) concentration of $n_{in} = 10^{18}\ cm^{-3}$. The low energy electrons (holes) become dark as they are captured by the donor (acceptor) BANDs while the higher energy electrons (holes) remain bright as they move freely and can recombine radiatively with each other. High energy bright electrons (holes) cannot recombine radiatively with low energy dark holes (electrons) because of the k-wavevector mismatch schematically shown in the lower panel of (c). A fraction of the charge carriers is eliminated by a recombination through non-radiative defects.

In a standard semiconductor the thermalized electrons and holes can recombine radiatively in a bimolecular process. In addition, they can recombine non-radiatively via deep defect centers and by the Auger recombination. A presence of positively and negatively charged BANDs adds a new very important modification to the recombination processes. As is schematically shown in Figure 1(c) the Coulomb potentials of BANDs separate the photoexcited carriers into two groups. The low kinetic energy electrons that are captured by the positively charged BAND donors are spatially separated from the low kinetic energy holes that are captured by negatively charged BAND acceptors. This reduces the radiative bimolecular recombination and the Auger recombination rates because the low energy, "dark" electrons do not encounter the low energy "dark" holes. In contrast, the carriers with large kinetic energy are not affected by the coulomb potentials and are uniformly spatially distributed. These high energy, "bright" electrons and holes can still recombine radiatively, contributing to photoluminescence, and can also participate in the Auger recombination. However, it should be emphasized that in addition to the, discussed above, real space separation of the dark electrons from the dark holes, the high energy large wavevector bright carriers are separated from the low energy small wavevector dark carriers in the k-space. Therefore, although the bright electrons (holes) can penetrate the regions occupied by dark holes (electrons) they cannot recombine radiatively with them because of a k-wavevector mismatch.

To provide a quantitative description of these processes we consider a HOIP with $2N_d$ of uniformly distributed BANDs. The average volume associated with each defect is $V_d = (2N_d)^{-1}$. A Coulomb potential of a donor (acceptor) will capture an electron (hole) when the attractive electrostatic potential energy is larger than the kinetic energy, E, of the carrier, $W_c(r) \geq E$

where

$$W_c(r) = \frac{e^2}{4\pi\varepsilon_0\epsilon_0 r} \tag{1}$$

Here $e$ is the electron charge, $\epsilon_0$ is the vacuum permittivity and $\varepsilon_0$ is the low frequency dielectric constant equal 18 in MAPbI$_3$[30]. The capture radius for a charge carrier with energy $E$ is given by,

$$r_c(E) = \frac{e^2}{4\pi\varepsilon_0\epsilon_0 E} \tag{2}$$

with the corresponding capture volume,

$$V_c(E) = \frac{4}{3}\pi(r_c(E))^3 \tag{3}$$

The fraction of the volume outside the Coulomb capture regions is where electrons and holes can encounter each other and recombine radiatively, producing PL. This "bright" fraction of free carriers is given by,

$$\begin{cases} V_{br}(E) = 1 - 2DN_d/(E)^3 & \text{for } 2DN_d/(E)^3 \leq 1 \\ V_{br}(E) = 0 & \text{for } 2DN_d/(E)^3 > 1 \end{cases} \tag{4}$$

where $D = \frac{4}{3}\pi \left(\frac{e^2}{4\pi\varepsilon_0\epsilon_0}\right)^3$.

The spatial separation of the low energy electrons captured by donors from low energy holes captured by acceptors prevents them from recombining radiatively. The fraction of these "dark" electrons and holes as a function of energy is equal to,

$$\begin{cases} V_{dr}(E) = 2DN_d/(E)^3 & \text{for } 2DN_d/(E)^3 \leq 1 \\ V_{dr}(E) = 1 & \text{for } 2DN_d/(E)^3 > 1 \end{cases} \quad (5)$$

It is important to emphasize that the capture of the dark electrons and holes by the Coulomb potentials is different from previously considered trapping process[31] in which electrons and/or holes are bound and immobilized by trapping Coulomb centers and do not contribute to the charge transport. In the capture process considered here the dark electrons and holes can still move around and transport charge, although their paths are separated as the electrons tend to reside around positively charged donors and holes around negatively charged acceptors. The bright and dark carriers are in thermal equilibrium with the lattice via emission and absorption of phonons. In addition, as will be discussed later, the carrier separation into bright and dark fractions modifies the non-radiative recombination rates occurring via deep level recombination centers and by Auger recombination.

To calculate the equilibrium fractions of the bright and dark electrons and holes we assume that the conduction and the valence band have the same parabolic dispersions with the effective mass $m_c^* = m_v^* = 0.2m_e$ [9], where $m_e$ is the free electron mass. An important and consequential feature of the model is the transformation between BANDs in the donor and acceptor configuration represented by the defect reactions[12]

$$2e_{PVSK} + 2D^+ \rightarrow D^0 + D^+ + e_{PVSK} \quad (6a)$$

$$D^0 + D^+ + e_{PVSK} \rightarrow A^0 + D^+ + e_{PVSK} \tag{6b}$$

$$A^0 + D^+ + e_{PVSK} \rightarrow A^- + D^+ + e_{PVSK} + h_{PVSK} \tag{6c}$$

which assure that the free conduction band electron ($e_{PVSK}$) and the valence band hole ($h_{PVSK}$) concentrations are balanced and equal. Consequently, although the following calculations are done for electrons in the conduction band, the same considerations apply to the holes in the valence band as well. Thus, unless specified, in the following analysis we will use the term electrons or carriers for both electrons in the conduction band and holes in the valence band.

The distribution of bright and dark electrons in the conduction band is given by,

$$n_{br}(E) = N_c V_{br}(E) E^{\frac{1}{2}} f_n(E) \tag{7}$$

and

$$n_{dr}(E) = N_c V_{dr}(E) E^{\frac{1}{2}} f_n(E) \tag{8}$$

where $f_n(E)$ is the Fermi-Dirac distribution function, $V_{br}$ and $V_{dr}$ are given by Eq. (4) and Eq. (5), respectively, and the effective density of states at the conduction band minimum is,

$$N_c = \frac{1}{2\pi^2} \left[\frac{2m^*}{\hbar^2}\right]^{3/2} \tag{9}$$

Under steady state illumination conditions, the incident light sustains a total electron concentration $n_{in}$ in the conduction band and the same concentration of holes in the valence band. This corresponds to the reduced quasi-Fermi energy, $\eta_{Fn} = E_{Fn}/k_B T$ which can be obtained from the following equation:

$$n_{in} = N_c(k_BT)^{3/2} \int_0^\infty \frac{z^{1/2}}{1+exp(z-\eta_{Fn})} dz = N_{cg} F_{1/2}(\eta_{Fn}) \qquad (10)$$

where $z = E/k_BT$, $N_{cg} = N_c(k_BT)^{3/2}$ and $F_{1/2}(\eta_{Fn})$ is the Fermi-Dirac integral of the order ½.

The energy distributions of bright and dark photoexcited electrons are then represented by Eq. (7) and (8), respectively, with the electron quasi-Fermi energy determined from Eq. (10). The hole quasi-Fermi energy is given by an analogous set of equations derived for hole statistics in the valence band. Figure 1(c) shows the thermal distribution for the photoexcited carrier concentration of $n_{in} = p_{in} = 2 \times 10^{18}\ cm^{-3}$ corresponding to the quasi-Fermi energies of electrons (holes) located at the conduction (valence) band edge.

3. Effects of BANDs on optical properties of perovskites

3.1 Optical absorption

For parabolic conduction and the valence bands the spectral dependence of the absorption is given by,

$$\alpha(h\upsilon) = \alpha_0 (h\upsilon - E_g)^{1/2} \qquad (11)$$

where $\alpha_0$ is a constant related to the strength of the optical coupling between the valence and the conduction band states. Although the light absorption is not directly affected by BANDs, they can have an indirect effect since, as will be shown later, they change the recombination rates and thus also the concentration of the photoexcited carriers. This has an effect on the optical absorption that depends on the occupation of the conduction by electrons and the valence band by holes. The sensitivity of the optical absorption to the band occupancy, or photobleaching effect, allows the study of the dynamics of the photoexcited carriers by measuring time dependence of the photoinduced transmission[32, 33]. The phenomenon is described in terms of the absorption coefficient modified by the band occupation,

$$\alpha_m(h\upsilon) = \alpha(h\upsilon)\left(1 - f_n[(h\upsilon - E_g)/2k_BT]\right)^2 \qquad (12)$$

where $f_n(E)$ is the Fermi-Dirac distribution function for electrons in the conduction band and holes in the valence band.

3.1 Charge carrier recombination and capture

The temporal evolution of the distribution of photoexcited charge carriers is governed by equations that account for the carrier generation as well as all possible routes for charge recombination and capture. Thus, in the present case, in addition to the standard processes that include: bimolecular radiative, Auger non-radiative and defect facilitated non-radiative recombination[27, 34, 35], we need to consider the process of capturing of electrons (holes) by donor (acceptor) BANDs that divides them into bright ($n_{br}$) and dark ($n_{dr}$) electrons (holes).

Capture and de-capture of bright electrons (holes) by BAND donors (acceptors) can be described by first order reactions:

Capture by Coulomb potential, generating dark carriers

$$n_{br} + (N_d - n_{dr}) \xrightarrow{k_{cp}} n_{dr} \qquad (13)$$

with the time dependence of their concentrations given by the equation,

$$\frac{dn_{br}}{dt} = -k_{cp}n_{br}(N_d - n_{dr}) \tag{14}$$

De-capture, generating bright carriers

$$n_{br} + N_d \xleftarrow{k_{-cp}} n_{dr} \tag{15}$$

$$\frac{dn_{br}}{dt} = k_{-cp}n_{dr} \tag{16}$$

where $k_{cp}$ and $k_{-cp}$ are the constants describing the capture and de-capture rates, respectively. The time dependence of $n_{br}$ and $n_{dr}$ is given by the counterbalance between those two reactions as described by the equation,

$$\frac{dn_{br}}{dt} = -\frac{dn_{dr}}{dt} = -k_{cp}n_{br}(N_d - n_{dr}) + k_{-cp}n_{dr} \tag{17}$$

At equilibrium and in absence of any other recombination processes, $\frac{dn_{br}}{dt} = \frac{dn_{dr}}{dt} = 0$ and the Eq. (17) yields a relationship between the capture and de-capture constants,

$$k_{-cp} = \frac{k_{cp}n_{breq}(N_d - n_{dreq})}{n_{dreq}} \tag{18}$$

where $n_{breq}$ and $n_{dreq}$ are the equilibrium concentrations of the bright and dark electrons, respectively. Using Eqs. (7) and (8) the ratio of bright to dark carrier concentrations can be expressed in terms of the equilibrium Fermi energy,

$$B_{eq} = \frac{n_{breq}}{n_{dreq}} = \frac{\int_0^\infty f_n(z)z^{1/2}V_{br}(z)dz}{\int_0^\infty f_n(z)z^{1/2}V_{dr}(z)dz} \tag{19}$$

and the Eq. (14) takes the form,

$$\frac{dn_{br}}{dt} = -k_{cp}(n_{br} - B_{eq}n_{dr})(N_d - n_{dr}) \tag{20}$$

The energetic separation of the photoexcited charge carriers into bright and dark fractions has a profound effect on the optical and charge transport properties of perovskites. Most notably, since the dark electrons are spatially separated from the dark holes, they cannot recombine in any processes that require close encounter of electrons and holes. Therefore, the radiative bimolecular and non-radiative Auger recombination processes are limited to the bright carriers only. In addition, the spatial localization of dark electrons and holes reduces the sample volume fraction they can penetrate which, in turn, reduces the probability of encountering non-radiative defect centers. Under the thermal quasi-equilibrium conditions with defined lattice and the carrier temperature the monomolecular non-radiative recombination for bright and dark electrons and holes are,

$$\frac{dn_{br}}{dt} = -\int_0^\infty R_{nr} N_{nr} N_{cg} V_{br}(z) z^{\frac{1}{2}} f_n(z) dz = -k_{nr} n_{br} \tag{21}$$

$$\frac{dn_{dr}}{dt} = -\int_0^\infty R_{nr} N_{nr} N_{cg} (V_{dr}(z))^2 z^{\frac{1}{2}} f_n(z) dz = -k_{nr} r_{dr} n_{dr} \tag{22}$$

where $k_{nr} = R_{nr} N_{nr}$ is the non-radiative recombination rate constant given by the product of the non-radiative recombination rate, $R_{nr}$ and the concentration of the non-radiative recombination centers, $N_{nr}$. The additional factor $V_{dr}(E)$ in Eq. (22) accounts for the restrictions on the penetration volume by the dark electrons (holes).

The effective non-radiative recombination rate of dark carriers is modified by the factor,

$$r_{dr} = \int_0^\infty (V_{dr}(z))^2 z^{\frac{1}{2}} f_n(z) dz / \int_0^\infty V_{dr}(z) z^{\frac{1}{2}} f_n(z) dz \tag{23}$$

Note that since $V_{dr}(E) \leq 1$ then also $r_{dr} \leq 1$ which means that in addition to being protected from radiative recombination the dark carriers have also the non-radiative recombination rate that is smaller than bright carriers. Also, the k-wavevector section rule eliminates the radiative recombination between the bright electrons (holes) and dark holes (electrons) because, as is schematically shown in Figure 1(c), and discussed in the supplementary information section there is a large difference of more than $3 \times 10^6 \ cm^{-1}$ between k-wavevectors corresponding to the average energy of the bright and dark carriers.

Combining all the recombination and capture processes and adding the carrier generation rate G yields two coupled equations for the time evolution of the photoexcited bright and dark electrons or holes,

$$\frac{dn_{br}}{dt} = -k_{cp}(n_{br} - B_{eq}n_{dr})(N_d - n_{dr}) - k_2 n_{br}^2 - k_3 n_{br}^3 - k_{nr}n_{br} + G \quad (24)$$

$$\frac{dn_{dr}}{dt} = k_{cp}(n_{br} - B_{eq}n_{dr})(N_d - n_{dr}) - k_{nr}r_{dr}n_{dr} \quad (25)$$

here $k_2$ and $k_3$ are the bimolecular and Auger recombination constants, respectively. It should be noted that $k_2$ which describes recombination of mobile high energy electrons and holes is expected to be larger than $k_{cp}$ describing capture of the slow, low energy electrons (holes) by immobile donors (acceptors).

At this stage it is important to highlight certain distinctive features of the dark charge carriers. Since the dark electrons (holes) reside mostly in the vicinity of BAND donors (acceptors) they are separated and cannot recombine radiatively. Also, dark electrons (holes) attracted by positively (negatively) charged donors (acceptors) are easily trapped and produce neutral BANDs that can transform between donor and acceptor configuration making them energetically indistinguishable from each other. Therefore, these neutral BANDs behave as a new class of quasiparticles, which in the following will be called bandons. They remain mobile, can store and transport energy of a photoexcitation and are protected from recombination. Since, the bandons are in thermal equilibrium with electrons and holes as shown by the reactions (6a) to (6c), they also provide a path for maintaining the balance between electrons and holes. Any local excess of electron concentration produces excess of bandons in donor configuration by neutralizing some donors (reaction (6a)). To reach equilibrium a fraction of excess bandons in the donor configuration transforms into acceptor configuration (reaction (6b)). Eventually the bandons in the acceptor configuration are ionized producing holes in the valence band (reaction (6c)). The process restores the balance between bandons in the acceptor and donor configurations and equalizes electron and hole concentrations. An analogous process occurs also for an excess hole case.

The partition of the carriers into the bright and dark fractions spatially separates the transport of the dark electrons and holes and is highly beneficial for the PV performance. The effect is somewhat similar to the separation of electrons and holes in interpenetrating phase-separated donor acceptor absorber blends that are used to reduce recombination and improve charge collection in polymer PVs[36, 37]. However, it is important to note a key difference between those two cases, as the spatial configuration of the blends is pre-determined during material preparation whereas the formation of the dark electron and dark hole transporting regions in HOIPs is a dynamic process in which the spatial distribution of the electrostatic field is affected by free carrier induced transformations between donor and acceptor configuration of randomly distributed BANDs. It can be also seen that the bandons' capability to store and transport the energy of optical excitation is expected to lead to effects similar to photon recycling that has been frequently invoked in explaining photophysical phenomena in HOIPs[27, 28, 38]. An evident and critical difference is that unlike isotropically emitted photons that can easily escape the sample the bandons are fully confined to the sample volume.

4. Model calculations of photophysical properties

In order to illustrate the effects of the BANDs on optoelectronic properties of perovskites we solve the Eqs. (24) and (25) for the cases corresponding to specific phenomena that can be determined experimentally. The purpose of the calculations is to give a general description of the main features of the model and to illustrate how it provides a consistent and unified explanation for a large variety of strikingly unusual and seemingly unrelated photophysical effects observed in HOIPs.

Time resolved and steady state measurements of the photoluminescence (PL), photoconductivity (PC) and photobleaching (PB) are the methods commonly used to discern properties and evaluate quality of perovskite materials[34, 35, 39, 40]. Standard interpretation of such experiments assumes that the photoexcitation produces electrons in the conduction band and holes in the valence band that can be treated as uniform free electron and hole gases. However, as is discussed above, this is not the case for HOIPs with BANDs because the photoexcited electrons and holes will be separated in two fractions with distinctly different characteristics. This, we will show, has a profound effect on the optoelectronic as well as charge transport and diffusion properties of HOIPs.

4.1 Time resolved photoeffects

The time resolved and the steady state PB, PL and PC are determined by the concentrations of the bright and dark electrons and holes which can be calculated using the coupled Eqs. (24) and (25). In our calculations the PL intensity depends on the concentration of the bright carriers and is equal to $k_2 n_{br}^2$ whereas PB depends on the occupation of the conduction and the valence band states and is related to the total concentration of the photoexcited carriers, $n_{tot} = n_{br} + n_{dr}$. The change in the optical absorption $\Delta A/A$ or transmission $\Delta T/T = -\Delta A/A$ can be calculated using Eqs. (10), (11) and (12), where the quasi-Fermi energies for electrons (holes) are determined from the total concentration of the photoexcited electrons (holes). It can be shown that for the nondegenerate carrier statistics, i.e. when the carrier concentration is smaller than about $10^{18}\ cm^{-3}$, the transmission change is just proportional to $n_{tot}$ and its spectral dependence is given by $\frac{\Delta T}{T}(h\nu) \sim n_{tot} \alpha(h\nu) Exp[(h\nu - E_g)/2k_B T]$. Finally, the PC is given by the contributions from the dark and bright electrons and holes to the conductivity and is proportional to $2(n_{br}\mu_{br} + n_{dr}\mu_{dr})$, where $\mu_{br}$ and $\mu_{dr}$ are the mobilities of the bright and dark electrons (holes), respectively. The mobility ratio $\mu_{dr}/\mu_{br}$ is rather difficult to estimate, although results of simple calculations presented in Figure S2 of the supplementary information section suggest that an approximate value of $\mu_{dr}/\mu_{br}$ should be equal to about 0.14. Therefore, the conductivity of the photoexcited carriers is assumed to be proportional to, $n_{br} + 0.14 n_{dr}$.

Currently there is no consensus on the values of the recombination constants in HOIPs. In our model calculations we have used, $k_2 = 8 \times 10^{-11}\ cm^3 s^{-1}$ for the bimolecular recombination constant, which falls into the broad range of the values for this parameter reported in the literature[41]. The carrier capture by BANDs depends on the mobility of the dark carriers and is expected to be smaller than $k_2$. Here we adopted the value $k_{cp} = 2 \times 10^{-11}\ cm^3 s^{-1}$. The third order Auger recombination process does not appear to be important at typical photoexcitation levels and as we will show later the laser cooling experiments provide an upper limit, $k_3 < 10^{-29}\ cm^6 s^{-1}$. In order to evaluate the effects of non-radiative defect trapping we have also added term with a recombination rate constant varying in the range $10^5\ s^{-1} \leq k_{nr} = R_{nr} N_{nr} \leq 10^7\ s^{-1}$.

Figure 2(a) and (b) show the calculated time dependencies of the total, $n_{tot}$ and the dark, $n_{dr}$ carrier concentrations for few different values of $N_d$ and initial concentrations of the photoexcited charge carriers, $n_0$. For $N_d = 0$ the total concentration shows a typical transient with the decay rate determined by a combination of the radiative bimolecular and a non-radiative monomolecular recombination. Adding BANDs results in a rapid capture of carriers and increase of the dark carrier concentrations until they reach the thermal equilibrium with the bright carriers. And since the dark carriers are protected from recombination this leads to a much slower decay of $n_{tot}$ at longer times. This effect is especially significant at larger $n_0$ where, as is seen in Figure 2(b), for $n_0 = 10^{18}\ cm^{-3}$ at 300 ns $n_{tot}$ is several times larger for $N_d = 5 \times 10^{18}\ cm^{-3}$ than for $N_d = 0$.

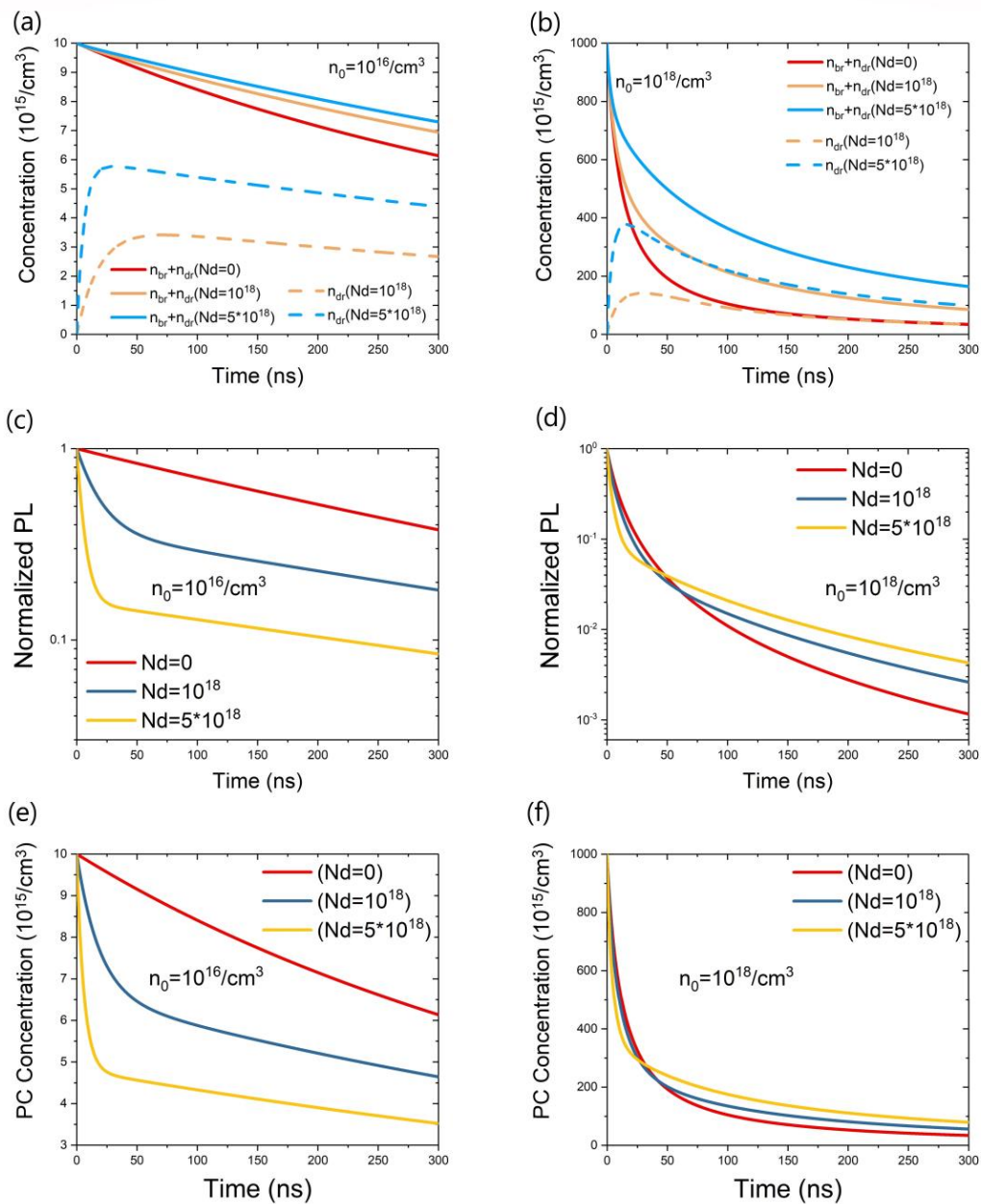

Figure 2. (a), (b) Time dependence of the total electron concentration (solid lines) and the dark electron concentration (dash lines) as well as corresponding photoluminescence (PL) (c), (d) and photoconductivity (PC) (e), (f) transients calculated for different BANDs concentrations, $N_d$, and two different initial concentrations of the photoexcited charge carriers, $n_0$. More PB, PL and PC transients at different time ranges corresponding to possible experimental conditions are presented in Figures S3 to S6 of the supplementary information section.

The time dependencies of $n_{tot}(t), n_{dr}(t)$ and $n_{br}(t) = n_{tot}(t) - n_{dr}(t)$ from Figure 2(a) and (b) can be used to calculate the transients of the photophysical properties such as PL, PB and PC that are commonly measured to evaluate properties and asses suitability of perovskite materials for PV applications. Since, for the nondegenerate carrier statistics the PB effect is proportional to the carrier concentration therefore, in relative terms, its transients are represented by $n_{tot}$ shown in Figure 2(a) and (b). Figure 2(c) and (d) show the calculated PL transients given by, $k_2 n_{br}^2$. For $N_d = 0$ the calculations show standard PL transients dominated by the monomolecular non-radiative recombination for the low excitation level, $n_0 = 10^{16}\ cm^{-3}$ (Figure 2(c)) and by the bimolecular radiative recombination at the higher excitation level, $n_0 = 10^{18}\ cm^{-3}$ (Figure 2(d)). A presence of BANDs strongly affects the PL transients even for the low excitation intensity where as it is seen in Figure 2(c) the PL intensity decays rapidly at short times with the decay rate increasing with increasing $N_d$. The effect is more pronounced for the higher photoexcitation case, $n_0 = 10^{18}\ cm^{-3}$ where, as is shown in Figure 2(d) the PL intensity decays by more than an order of magnitude in the first 20 ns for $N_d = 5 \times 10^{18}\ cm^{-3}$. It is important to note that the decay does not originate from a loss of carriers but is fully attributable to shown in Figure 2(b) rise of the concentration of dark carriers or bandons that are protected from the radiative recombination. The photoexcitation energy stored by dark carriers is eventually released at longer times when, as is illustrated in Fig. 2 (d) at 300 ns the PL intensity for $N_d = 5 \times 10^{18}\ cm^{-3}$ is few times large than for $N_d = 0$. Thus, it can be argued that the dark carriers (or bandons) serve as a medium capturing and storing the photoexcitation energy at short times and releasing it later when the dark carriers start to back transfer to maintain the thermal equilibrium and replenish the bright carrier reservoir depleted by the radiative recombination.

Photoluminescence transients with rapid initial and much slower decays at longer times have been routinely observed in perovskite films[34, 39]. At low and moderate photoexcitation levels the initial decay is typically attributed to non-radiative recombination processes such as deep level facilitated trapping of one type of carrier by defect centers followed by a release of the trapped carriers[42]. Overall, such a process always leads to a non-radiative removal of the photoexcited carriers and is detrimental for both light emission and charge transport. In contrast, although the capture of the photoexcited carriers by BANDs results in an initial fast reduction of PL it does not eliminate the carriers but protects them from recombination and shifts all the transient effects to longer times. At very high excitation levels the initial decays are often attributed to the Auger recombination which requires the assumption of unusually large recombination constants of about $10^{-28}\ cm^6 s^{-1}$ [27].

The calculated BANDs induced reduction of the recombination rates and, shown in Figure 2(b), the slow decay of the carrier concentration explains the experimentally observed long lived PL signals[31]. Also, it helps to understand the high-resolution micro-PL study that has shown large differences in the PL intensity and transients between different grains in polycrystalline perovskite films[17]. It was observed that the dark grains, with rapid initial PL transient exhibit a delayed PL that converges on the PL intensity of the bright grains that exhibit a faster decay at long times. This behavior can be easily explained assuming the dark grains with larger concentration of BANDs capture the carriers and, as is seen in Figure 2(c) and (d), delay the PL to longer times. However, it is important to emphasize that these spatial variations in the PL do not have any detrimental effect on the charge carrier transport or PV performance since, as is seen in Figure 2(a) and (b) the materials with larger $N_d$ (dark grains) actually have the $n_{tot}(t)$ larger than the materials with smaller $N_d$ (bright grains). Also, the variations of the $n_{tot}(t)$ between bright (small $N_d$) and dark (large $N_d$) grains are much smaller that the variations of the PL intensity. This explains why such optically inhomogeneous materials make so exceptionally good PVs.

4.2 Steady-State Illumination

In order to evaluate the effects of BANDs on the optical characteristics under steady-state or continuous-wave (cw) illumination conditions we solve Eqs. (24) and (25) with a spatially uniform, nonzero generation rate $G$. Figure 3(a) shows the calculated time dependence of the total and the dark electron concentrations for different concentrations of BANDs and the generation rate $G = 5 \times 10^{21} \ cm^{-3}s^{-1}$ roughly corresponding to the one sun illumination of $0.1 \ Wcm^{-2}$ switched on at $t = 0$. As expected, for $N_d = 0$ the carrier concentration increases with time and saturates at $n_{tot} = n_{br} \cong 4 \times 10^{15} \ cm^{-3}$. Adding BANDs leads to formation of bandons ($n_{dr}$) and an increase of $n_{tot}$. However as is seen in Figure 3(b) formation of bandons that are protected from radiative recombination results in a large reduction of the PL intensity. This is because in the presence of BANDs the $n_{dr}$ is increasing reducing the concentration of $n_{br}$ and thus also the PL intensity. At the saturation level the ratio of the bright and dark carrier concentrations is constant and equal to the thermal equilibrium ratio, $B_{eq}$ given by Eq. (19).

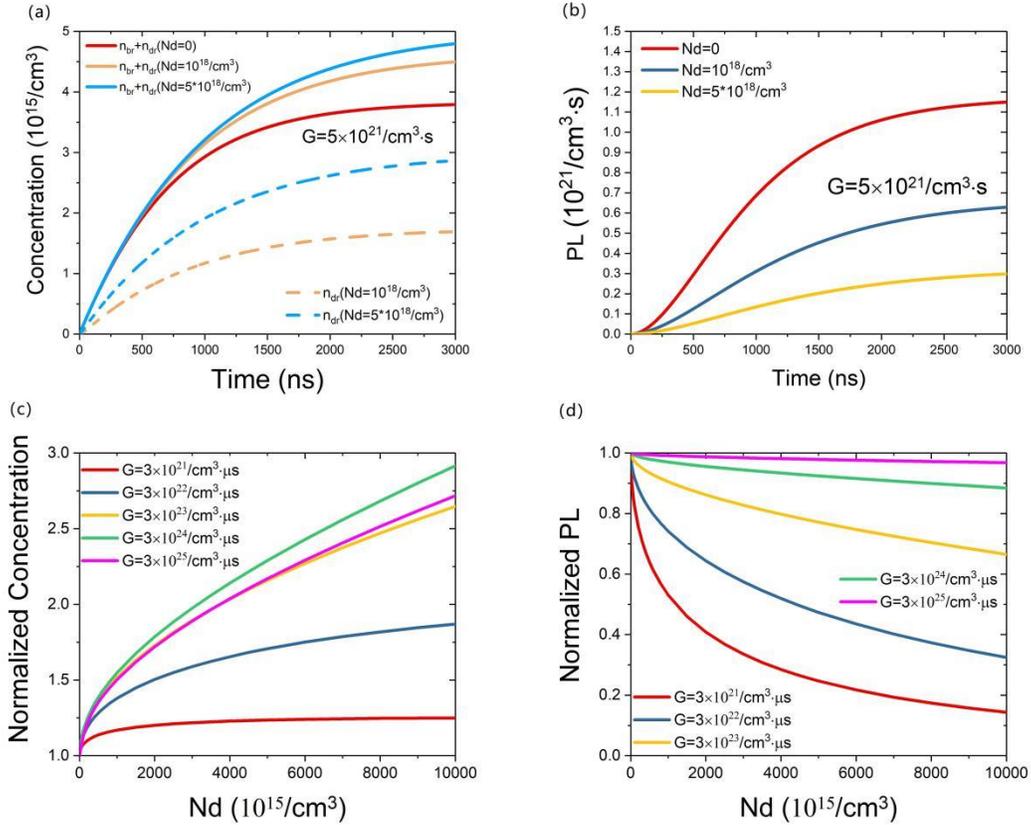

Figure 3. Temporal evolution of the total, ($n_{tot}$) and the dark, ($n_{dr}$) carrier concentrations (a) as well as PL intensity (b) at the one-sun equivalent steady-state illumination, $5 \times 10^{21}\ cm^{-3} s^{-1}$ for different concentrations of BANDs, ($N_d$). Calculated dependence of the saturated (steady-state) total carrier concentration ($n_{tot}$) (c) and the PL intensity (d)

on the BANDs concentration. Both $n_{tot}$ and PL are normalized to their values at $N_d = 0$. The rise (a) and the saturation level (c) of the total concentration is increasing with $N_d$ whereas, in a stark contrast, the rate of the initial rise (b) and the steady state saturation levels (d) of the PL intensity are decreasing with increasing $N_d$.

To gain a better insight into how the BANDs affect the photophysical properties under steady state illumination conditions and how they relate to the device performance we have calculated the dependencies of the saturation values of $n_{tot}$ and PL on the BAND concentration. The results, normalized to the values at $N_d = 0$, are shown in Figure 3(c) and (d). It is seen in Figure 3(d) that for the illumination intensity $G = 3 \times 10^{21} \ cm^{-3}s^{-1}$ roughly corresponding to the one sun illumination the PL intensity is decreasing by a factor of about 5 when $N_d$ changes from 0 to $6 \times 10^{18} \ cm^{-3}$, at the same time Figure 3(c) shows that the total carrier concentration is increasing by about 30%. This explains a puzzling and apparently contradictory feature of thin polycrystalline HOIP films that are composed of grains with drastic variations of the PL intensity[17, 43] but still work as PV absorbers with superior uniform charge transport properties. Also, they demonstrate that the conventional approach relating performance of PV devices to the intensity and uniformity of the PL signal does not apply to HOIPs. An interesting feature of the results presented in Figure 3(d) is that at high carrier generation rates $G$, the PL intensity becomes insensitive to the concentration of BANDs. This is in excellent agreement with the observation that large spatial variations of the PL intensity under low illumination are washed out at high illumination intensities[17].

4.3 Diffusion of photoexcited carriers

The above calculations show that the BANDs induced separation of the photoexcited carriers into bright and dark fractions greatly affects the carrier dynamics and the photophysical characteristics of HOIPs. Therefore, it is also expected to impact the photoexcited carrier diffusion which plays a critical role in determining the performance of PV devices. In standard semiconductors the electron mobility is typically significantly larger than the hole mobility. This, in addition to most likely different electron and hole non-radiative recombination rates leads to a large difference in the diffusion lengths resulting in an unbalanced transport and charge accumulation effects. In many instances this is an important factor affecting the performance of PV devices.

Measurements of the charge diffusion have shown long and balanced diffusion with diffusion lengths of more than 1 micron for electrons and holes in perovskite thin films[15, 16]. Remarkably, the large diffusion lengths have been observed in nonuniform polycrystalline films with large spatial variations of PL intensity that appeared to indicate a highly nonuniform distribution of non-radiative recombination centers. Even more unexpectedly it was shown that photoexcited electrons and holes can diffuse and produce detectable PL and photocurrent tens of microns away from the excitation point[15, 38]. It was argued that such long-distance transport of the photoexcitation can be explained by an efficient photon recycling effect[38]. Here we present an alternative and, in our opinion more plausible, explanation based on the BANDs induced charge separation and formation of bandons that are protected from recombination and can store and transport the photoexcitation energy over long distances.

In order to illustrate and understand how the capture of charge carriers by BANDs and formation of bandons affect the carrier diffusion we perform model calculations for a material system simulating properties of HOIPs. In addition, we assume that electrons (n) and holes (p) have different diffusion constants. In the calculations we consider a simple case of one-dimensional diffusion where electron-hole pairs are photoexcited in the plane, $x = 0$ with a constant generation rate, $G$. The diffusion profiles of the bright and dark electrons and holes are given by solutions of four coupled differential equations:

$$\frac{\partial n_{br}}{\partial t} = G + D_{br,n} \frac{\partial^2 n_{br}}{\partial x^2} - k_{cp}(n_{br} - B_{eq}n_{dr})(N_d - n_{dr}) - k_2 n_{br}p_{br} - k_{nr}n_{br} \quad (26)$$

$$\frac{\partial n_{dr}}{\partial t} = D_{dr,n} \frac{\partial^2 n_{dr}}{\partial x^2} + k_{cp}(n_{br} - B_{eq}n_{dr})(N_d - n_{dr}) - k_{nr}r_{dr}n_{dr} - k_0 N_d(n_{dr} - p_{dr}) \quad (27)$$

$$\frac{\partial p_{br}}{\partial t} = G + D_{br,p} \frac{\partial^2 p_{br}}{\partial x^2} - k_{cp}(p_{br} - B_{eq}p_{dr})(N_d - p_{dr}) - k_2 n_{br}p_{br} - k_{nr}p_{br} \quad (28)$$

$$\frac{\partial p_{dr}}{\partial t} = D_{dr,p} \frac{\partial^2 p_{dr}}{\partial x^2} + k_{cp}(p_{br} - B_{eq}p_{dr})(N_d - p_{dr}) - k_{nr}r_{dr}p_{dr} + k_0 N_d(n_{dr} - p_{dr}) \quad (29)$$

where $D_{i,j}$ with $i = br, dr$ and $j = n, p$ are the diffusion constants for bright and dark electrons and holes and the other symbols have the same meaning as in Eqs. (24) and (25). The additional term $k_0 N_d(n_{dr} - p_{dr})$ in Eqs. (27) and (29) describes the transition of bandons between donor and acceptor configuration that maintains the balance between the concentration of electrons and holes as described by the reactions 6(a) to 6(c)[12].

The second order partial differential equations (26) to (29) were solved numerically using the finite difference method with codes written in Python 3. In addition to previously used values, $k_{cp} = 2 \times 10^{-11}\ s^{-1}$, $k_2 = 8 \times 10^{-11}\ cm^3 s^{-1}$ we assume that the transformation rate between BANDs in the donor and acceptor configuration is given by a constant $k_0 = 1 \times 10^5\ s^{-1}$. Using the longitudinal optical phonon frequency $v_{opt} = 2.8 \times 10^{12}\ s^{-1}$ [44] as an attempt frequency, this value of $k_0$ corresponds to about 0.43 eV energy barrier for the transformation between the donor and acceptor configuration of BANDs.

To illustrate how the BANDs facilitate equilibrium between electrons and holes leading to balanced carrier diffusion we assume that the electron diffusion constant is 10 times larger than the diffusion constant of holes (this is a typical ratio for standard compound semiconductors). In a customary consideration of an unbalanced bipolar diffusion problem the coulomb attraction between faster diffusing electrons and slower diffusing holes leads to a common, effective diffusion constant $D_b = 2\frac{D_e D_h}{D_e + D_h}$, where $D_e$ and $D_h$ are, respectively, the diffusion constants of electrons and holes[45]. Note that, as expected, for a large diffusion constant ratio the bipolar diffusion rate is mainly determined by the slower diffusing carriers.

Figure 4 shows steady state 1D diffusion profiles calculated for the electron-hole pair generation rate of $G = 10^{26}\ cm^{-3} s^{-1}$ at $x = 0$. To demonstrate the protective role of BANDs the calculations with $N_d = 0$ and $N_d = 10^{18}\ cm^{-3}$ were performed for a wide range of the non-radiative recombination constants $10^5 \leq k_{nr} \leq 10^7\ s^{-1}$. The results in Figure 4 demonstrate that the introduction of BANDs extends the diffusion resulting in higher total concentration of the carriers (Figure 4(a)) as well as the PL intensity (Figure 4(b)) at long distances. The effect is most pronounced for higher values of $k_{nr}$ and, as is seen in Figs. 4(a) and 4(b) for $k_{nr} = 10^7\ s^{-1}$, adding $N_d = 10^{18}\ cm^{-3}$ of BANDs results in more than 3 orders of magnitude increase in $n_{tot}$ and more than 6 orders of magnitude increase in the PL intensity at the distance $x = 50\ \mu m$.

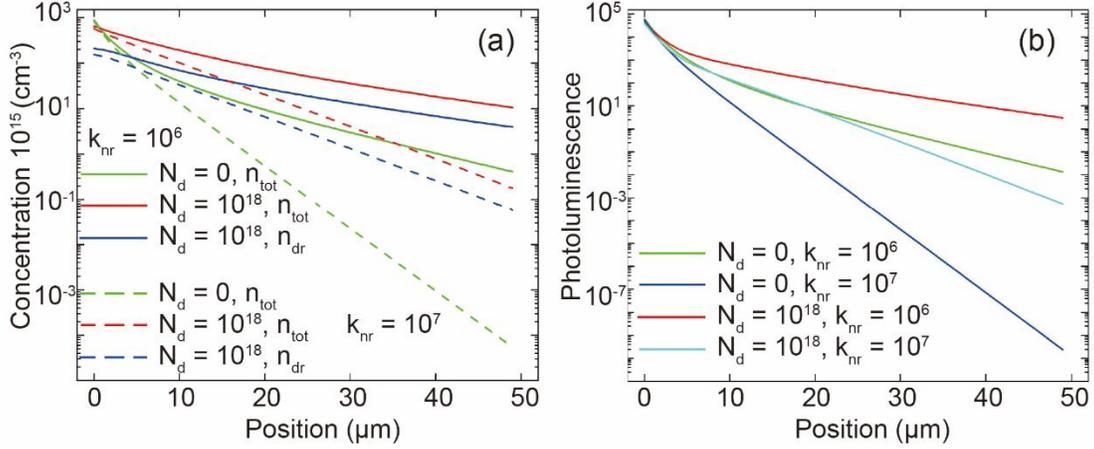

Figure 4. Calculated dependence of the steady state diffusion profiles on the non-radiative recombination rate, $k_{nr}$ and the concentration of BANDs, $N_d$ for $N_d = 0\ and\ N_d = 10^{18}\ cm^{-3}$. The total and the dark carrier concentration (a) and PL intensity (b) for $k_{nr} = 10^6$ and $10^7\ s^{-1}$. A generation rate of $G = 10^{26}\ cm^{-3} s^{-1}$ at $x = 0$ was assumed.

The efficient, long-distance diffusion of the bandons combined with shown in Figure 3(c) storage of the photoexcited carriers have interesting consequences for the photophysical properties of polycrystalline HOIPs with spatial variations of stoichiometry and nonuniform distribution of BANDs. The results in Figure 3(a) and (b) indicate that the dark grains with larger $N_d$ exhibit less intense PL but have larger total carrier concentrations whereas bright grains with smaller $N_d$ have stronger PL but smaller total carrier concentration. The difference in the total carrier concentration originates from higher concentration of bandons in the grains with larger $N_d$. Under steady state illumination conditions, the bandons or dark carriers will diffuse from the high $N_d$ (dark) to lower $N_d$ (bright) grains where they can become bright and recombine radiatively. This leads to an unusual effect in which dark grains act as a source supplying the carriers and enhancing PL in the bright grains. It is also in a stark contrast to standard polycrystalline semiconductors where the dark grains, with a strong non-radiative recombination, act as sinks depleting the concentration of carriers in the bright grains[46].

Another important feature of the bandons facilitated diffusion is the evolution of the PL spectra with the distance from the excitation source. The spectral dependence of the PL is proportional to the square of the energy distribution of the bright carrier concentration:

$$PL(hv) \sim \left[ V_{br}(E) E^{\frac{1}{2}} f_n(E/k_B T) \right]^2 \tag{30}$$

where $E = \frac{hv - E_g(T)}{2}$, $V_{br}(E)$ is given by Eq. (4) and $f_n(z)$ is the Fermi-Dirac distribution function. The calculated color maps of the PL spectra corresponding to the diffusion profiles presented in Figure 4 are shown in Figure 5(a) and (b). The map in Figure 5(a) shows that, as expected for the $N_d = 0$ case, the PL intensity is decreasing with increasing distance without any significant change in the spectral distribution of the emitted photons. In a stark contrast, as is seen in Figure 5(b), $N_d = 10^{18}\ cm^{-3}$ of BANDs not only extends the PL to larger distances but also results in a distinct blue shift of the spectra. This anti-Stokes shift will be later discussed in the context of the laser cooling effects. Here we only note that the value of the blue shift is decreasing with the distance in a manner similar to the experimentally observed energy shift in the PL spectra measured as a function of distance from the excitation in perovskite thin films[38].

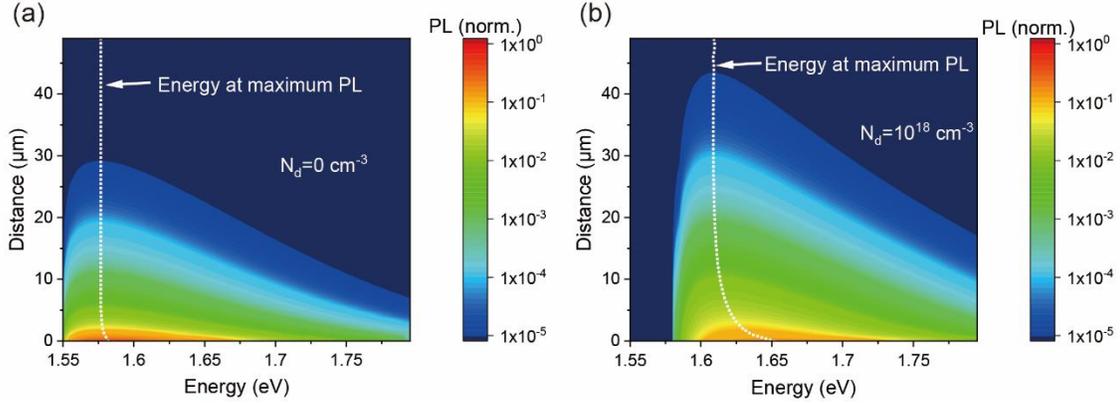

Figure 5. Color maps of the distance dependence of the PL spectra of diffusing carriers for (a) $N_d = 0$ and (b) $N_d = 10^{18}\ cm^{-3}$. The white dashed lines show the distance dependence of the energy of the PL maximum; when BANDs are present, the blue-shifted peak of PL spectrum exhibits a much larger red-shift as the distance away from the excitation source is increased.

The results presented in Figures 4 and 5 show that BANDs have a very large effect on the carrier diffusion extending the carrier concentration and PL profiles to longer distances. Although here we consider a simple one-dimensional, plane excitation geometry, the results are directly relevant to experimental observations that were performed in a point excitation geometry[38]. Previously the results of the long-distance diffusion[38], as well as, PL and conductivity transients[27] were explained by a photon recycling which can be viewed as a form of storage of the photoexcitation energy within the sample. It was shown that the photon recycling can be formally accounted for by reducing the bimolecular radiative recombination rate $k_2$[27] by the factor equal to the probability of the photon escape, $P_{esc}$ which has been estimated to be equal to about $0.12$[27, 47]. Calculations in Figures 6(a) and (b) show that, as expected, the eight-fold reduction of $k_2$ for $N_d = 0$ expands the diffusion profiles resulting in increases of the $n_{tot}$ by the factor of 7 and the PL intensity by the factor of 14 at the distance of $50~\mu m$. However, it is also seen in these figures that retaining the same value of $k_2 = 8 \times 10^{-11}~cm^3 s^{-1}$ and adding $N_d = 10^{18}~cm^{-3}$ of BANDs leads to an even larger extension of the diffusion profiles, with a factor of about 15 increase of the $n_{tot}$ and a very large increase by the factor of 200 for the PL intensity at the same distance of $50~\mu m$. This demonstrates that the bandons-enabled storage and transfer of the photoexcitation energy can very well account for the long-distance diffusion of carriers without invoking photon recycling.

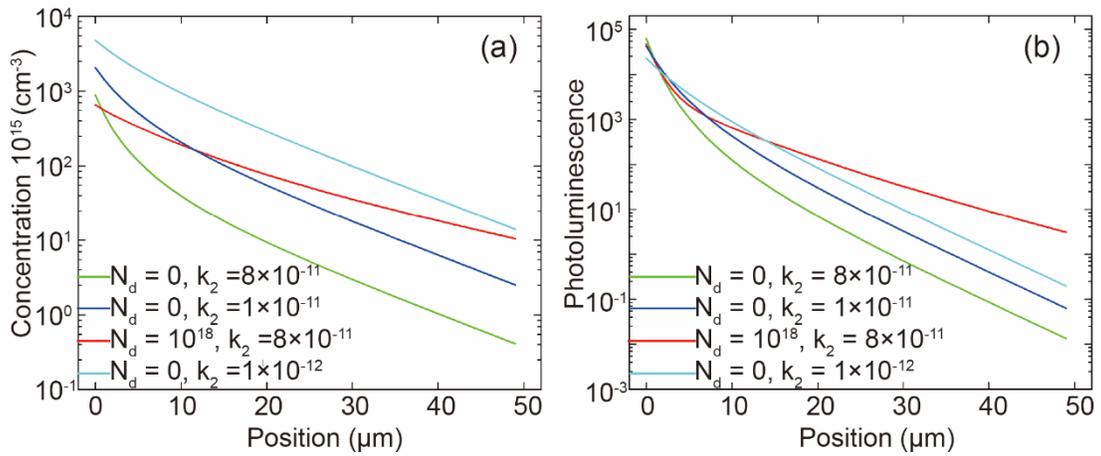

Figure 6. Effect of the bimolecular radiative recombination rate, $k_2$ on the diffusion profiles of the total carrier concentration (a) and the PL intensity (b) for $N_d = 0$ and $N_d = 10^{18}~cm^{-3}$. A generation rate of $G = 10^{26}~cm^{-3}s^{-1}$ at $x = 0$ was assumed.

5. Anti-Stokes Effect

The above considerations show that the BANDs induced separation of photoexcited charge carriers into bright and dark electrons and holes and formation of bandons have a profound effect on a large variety of photophysical properties of perovskites that depend on the photocarrier recombination processes. In this section we examine how the energy distributions for bright and dark carriers affect the relationship between absorption and emission spectra and provide explanation for some unusual properties that appear to be specific to HOIPs.

A shift of the PL emission to an energy higher than the photoexcitation energy is known as an anti-Stokes effect. This unique and rare phenomenon requires demanding conditions that are difficult to satisfy in standard semiconductors[48]. Here we show that an anti-Stokes shift of the PL spectrum occurs in HOIPs with the concentration of BANDs large enough to induce significant separation of the photoexcited carriers into bright and dark fractions. We will demonstrate that it provides a straightforward explanation for a number of unique phenomena observed in perovskites including: a direct observations of the PL anti-Stokes shift[19, 22], the energy difference between surface and bulk PL of single crystals[20, 21], and the optical cooling effects[22, 49].

As is illustrated in Figure 7(a) the threshold for photon absorption associated with electron transfer from an occupied state in the valence band to an unoccupied state in the conduction band is defined by the bandgap, $E_g$. In the presence of BANDs the absorption edge is lower than the threshold for the photon emission that can happen only when, as is shown in Figure 7(a), a bright electron in the conduction band recombines with a bright hole in the valence band. The energy difference is provided by lattice phonons that are absorbed to maintain thermal equilibrium between dark and bright carriers.

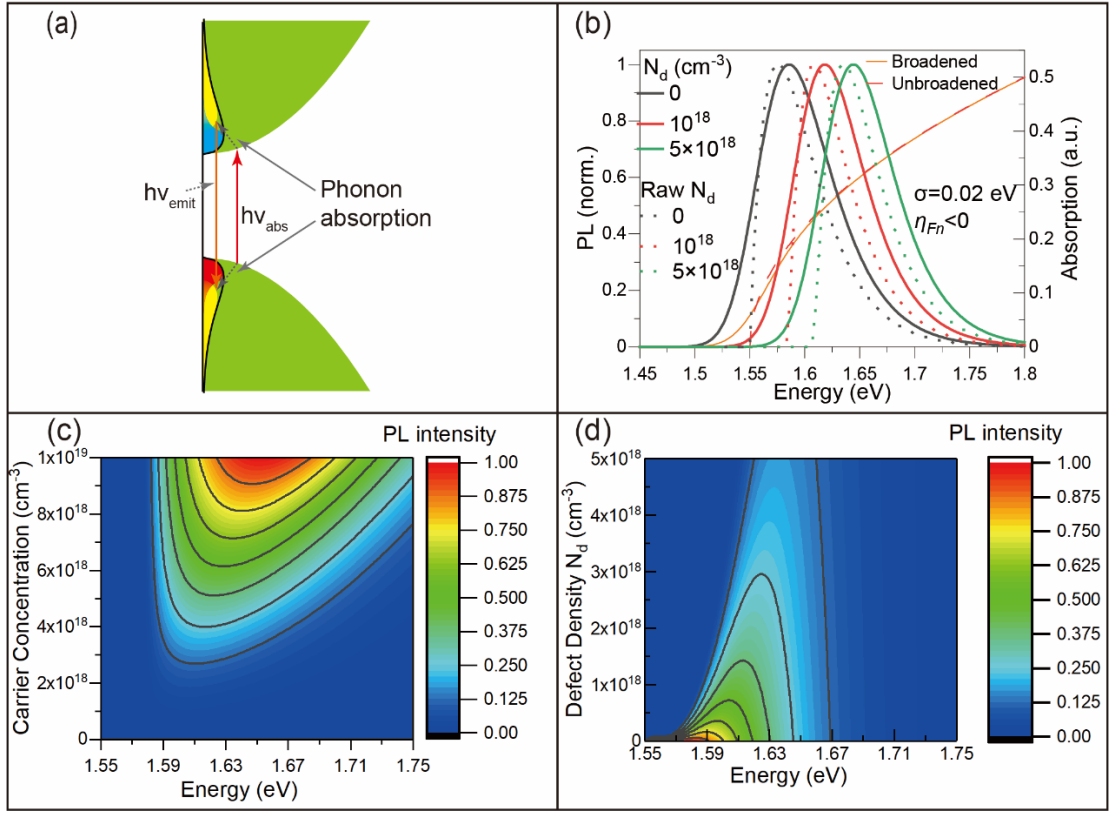

Figure 7. Anti-Stokes effect in HOIPs, (a) schematic illustration describing the mechanism of the anti-Stokes effect (b) calculated PL spectra for different concentrations of BANDs (left hand axis) along with the absorption spectrum (right hand axis). The inhomogeneously broadened curves are represented by solid lines. (c) color map of the PL intensity spectra vs. the concentration of the photoexcited electron-hole pairs, (d) color map of the PL intensity vs. concentration of BANDs for $n_{tot} = p_{tot} = 2 \times 10^{18}\ cm^{-3}$ corresponding to the quasi-Fermi energies for electrons and holes positioned at the band edges.

To calculate the anti-Stokes shift, i.e. the difference between the energy of emitted and absorbed photons, we again assume that there is a full symmetry between the distributions of photoexcited electrons and holes.

Photoluminescence spectra for different concentrations of BANDs calculated using Eq. (30) along with the absorption spectrum for $E_g = 1.55\ eV$ are plotted in Figure 7(b). The calculations were done for the nondegenerate carrier distribution case. The calculated PL spectra shift to higher energy with increasing concentration of BANDs and as is shown in Figure 7(b), an inhomogeneous broadening of the spectra results in a slight upward shift of the maximum PL energy. The PL spectrum for the $N_d = 0$ case shows the standard anti-Stokes shift $\Delta E_{AS}$ of about 30 meV resulting from a thermalization of the distribution of the photoexcited carriers. This small effect has been previously considered as a potential mechanism for a laser cooling in semiconductors[48]. However, as is seen in Figure 7(b) adding $N_d = 5 \times 10^{18}\ cm^{-3}$ of BANDs results in a large, three-fold increase of the anti-Stokes shift $\Delta E_{AS}$ to about 90 meV. As will be shown in the next section this large enhancement of the anti-Stokes shift is essential for understanding of the experimentally observed laser cooling effect in MAPbI$_3$[22].

The anti-Stokes shift does not depend on the carrier concentration for the nondegenerate carrier statistics, i.e. when $n_{tot} = p_{tot} < 2 \times 10^{18}\ cm^{-3}$ and the quasi-Fermi energies for photoexcited electrons and holes are in the bandgap. However, as is shown in Figure 7(c) the PL peak shifts to higher energy and the PL intensity is increasing with increasing carrier concentration for $n_{tot} = p_{tot} > 2 \times 10^{18}\ cm^{-3}$ when the electron (hole) quasi-Fermi energy enters the conduction (valence) band. The results in Figure 7(d) show that the PL peak shifts to higher energy but its intensity is decreasing with increasing concentration of BANDs.

5.1 Surface versus bulk photoluminescence in single crystals

The BANDs-induced anti-Stokes effect provides a clear-cut explanation for the observed difference in the PL spectra measured in thin polycrystalline films and bulk crystals[20, 21]. Detailed PL studies have demonstrated that the emission spectra of thick single crystals of HOIPs depend on whether the photoexcitation occurs close to the surface or in the bulk of the crystal[20, 21]. Interestingly and somewhat unexpectedly, excitation of a subsurface region with one, above the bandgap energy photon, produced a PL that was blue shifted relative to the PL photoexcited with two sub-bandgap photons and originating from the bulk of the crystal. The blue shifted subsurface PL spectra were found to be quite similar to the emission spectra of thin polycrystalline films[20, 21]. In addition, it was observed that the PL emission energy can be tuned between these two extrema by changing the ratio of the surface to bulk contribution using the photoexcitation with different photon energies[21] or single crystal platelets with different thicknesses[22].

In order to explain this effect, we note that, as was discussed before[12], the concentration of BANDs is related to local deviations of crystal stoichiometry that depend on the crystallinity and/or proximity to surfaces and interfaces. Therefore, higher concentrations of BANDs are expected in thin polycrystalline films rather than in single crystals. However, even in the case of single crystals significant deviations from stoichiometry and larger concentrations of BANDs can exist close to the surface. Figure 8(a) shows the PL spectra calculated using Eq. (30) along with the experimental PL data obtained with single-photon and two-photon excitations of a MAPbI3 crystal[20]. It is seen that the blue shift of the one-photon relative to the two-photon spectrum can be well explained assuming an average BAND concentration of $N_d = 2.5 \times 10^{18}\ cm^{-3}$ in the subsurface region and a much lower concentration of $N_d \leq 10^{15}\ cm^{-3}$ in the bulk of the crystal. The relative peak positions and the shapes of the spectra are reasonably well reproduced by the calculations that include a small inhomogeneous broadening. Also, the results in Figure 8(b) demonstrate that the same set of parameters can explain experimentally observed large differences in the time dependence of the PL intensity from the subsurface and bulk region[20] confirming that both effects, the blue shift of the PL energy and the faster initial PL decay originate from a higher concentration of BANDs and an increased capture of the photoexcited carriers in the subsurface region.

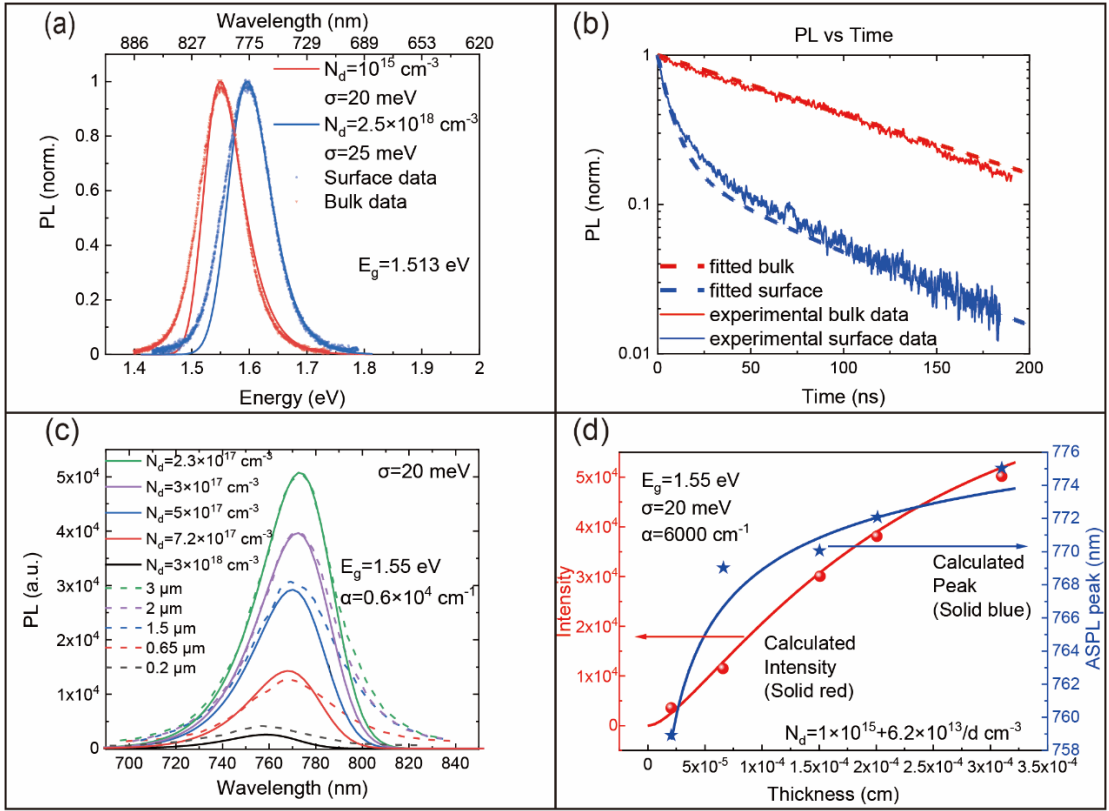

Figure 8. Effects of the anti-Stokes shift on the PL in single crystals. (a) comparison of the experimental and calculated PL steady state spectra for the bulk, two photon excitation (red) and for the subsurface, one photon excitation (blue) conditions. The calculated spectra are represented by the solid lines. (b) The PL transients for the bulk (red) and the subsurface (blue) regions calculated using the BANDs concentration obtained from the fits in (a) and the monomolecular non-radiative recombination constant $k_{\mathrm{nr}} = 4.5 \times 10^5\ s^{-1}$. The experimental results in (a) and (b) were adopted from ref. 20. (c) Determination of the average BANDs concentration from comparison of the calculated and measured PL spectra for single crystal platelets of different thickness. (d) Thickness dependence of the anti-Stokes shifted PL peak energy and the PL maximum intensity. The BANDs concentration $N_d(d)$ was obtained by averaging over the subsurface and the bulk regions of the platelets.

(for more details see the supplementary information section). The experimental results in (c) and (d) were adopted from ref[22].

Further support for the BANDs based interpretation of the PL energy shift has been provided by a systematic PL study of MAPbI$_3$ single crystal platelets, which has shown that PL peak energy is decreasing whereas its intensity is increasing with increasing platelet thickness. As is seen in Figure 8(c) and (d) these dependencies are again well explained by calculations assuming larger concentration of BANDs in the subsurface region compared with the bulk of the platelets. Further details of the calculations are presented in the supplementary information section.

5.2 Optical or Laser Cooling

The most intriguing and potentially most consequential effect of the large BANDs-induced anti-Stokes shift is an optical or laser cooling effect. It is an effect that requires a very efficient up-conversion of low energy absorbed photons into higher energy emitted photons[48]. Remarkably, a large and clearly observable optical cooling has been observed in platelets of MAPbI$_3$ synthesized on mica substrates[22]. It is shown here that the experimentally measured cooling can be well understood in terms of the BANDs induced anti-Stokes shift discussed above.

To calculate the laser cooling effect, we consider a perovskite platelet illuminated by a laser with a photon energy, $h\upsilon_L$ and the incident photon power flux $I_0$. The lattice temperature change, $\Delta T$ is proportional to the difference in the power density of absorbed and emitted light,

$$\Delta P_{dep} = G(h\nu_L)(h\nu_L - r(h\nu_L)h\nu_{emit}) \qquad (31)$$

here $h\nu_{emit}$ is the average energy of emitted photons and $0 < r(h\nu_L) \leq 1$ is the external PL efficiency that accounts for the crystal heating by the photon energy deposited in the sample through non-radiative recombination processes. The carrier generation rate is given by,

$$G(h\nu_L) = \frac{I_0(1-exp[-\alpha_m(h\nu_L)d_p])}{d_p} = P_L(h\nu_L)/h\nu_L \qquad (32)$$

where the modified absorption coefficient, $\alpha_m(h\nu_L)$ given by Eq. (12) accounts for possible absorption bleaching effects expected at high excitation levels when the quasi-Fermi energies move into the bands, and $P_L(h\nu_L)$ is the excitation laser power density absorbed in the sample. As is seen from Eq. (32) the carrier generation rate and thus also the power density deposited in the sample depends on the depth of the photoexcited region under the illuminated surface. In the cooling experiment[22] the temperature change was measured with PL probe excited with the photon energy of $h\nu_p = 1.85\ eV$ yielding an approximate value of $d_p = 1/\alpha_m(h\nu_p) \cong 230\ nm$ for the thickness of the probed region. The cooling efficiency is then given by,

$$\frac{\Delta P_{dep}}{P} = 1 - r(h\nu_L)h\nu_{emit}/h\nu_L \qquad (33)$$

The average energy of emitted photons can be calculated from,

$$h\nu_{emit} = \frac{\int_0^\infty (2E+E_g)PL(E)dE}{\int_0^\infty PL(E)dE} \qquad (34)$$

where $PL(E)$ is the spectral distribution of the photoluminescence of the bright carriers given by Eq. (30).

The external PL efficiency defined as,

$$r(hv_L) = k_2 n_{br}^2 / G(hv_L) \qquad (35)$$

is determined by the steady state concentrations of photoexcited bright carriers that were obtained solving Eqs. (24) and (25). In the calculations of the spectral dependence of the laser cooling we have adopted the values of the bimolecular recombination, $k_2 = 8 \times 10^{-11}\ cm^3 s^{-1}$, and the BANDs capture, $k_{cp} = 2 \times 10^{-11}\ cm^3 s^{-1}$ constants, the same as those used for the calculations of transients and diffusion processes in the preceding sections. The cooling efficiency is strongly dependent on the PL efficiency $r(hv_L)$, that is determined by non-radiative recombination processes. Here we consider the monomolecular non-radiative recombination via defect centers determined by the recombination constant, $k_{nr}$ that plays a more significant role at low carrier concentrations and the Auger recombination given by the constant, $k_3$ that is more important at high carrier densities.

Figure 9(a) shows the dependencies of the photon energy corresponding to the maximum of the PL intensity and the average emitted photon energy on the concentration of BANDs. As expected, adding BANDs results in a rapid increase of both energies. The average energy of emitted photons increases from 1.60 eV at $N_d = 0$ to 1.66 eV for $N_d = 5 \times 10^{18}\ cm^{-3}$. It is seen that under these idealized assumptions even for $N_d = 0$ the average energy of the emitted photons is larger than the band gap energy. This standard photon up-conversion effect[48] has been previously considered as a potential mechanism for the laser cooling in semiconductors. However, it was found that a very high, close to 100% external PL efficiency is required to explain laser cooling in HOIP platelets[22]. Here we show that the BANDs-induced increase of the average energy of the emitted photons greatly relaxes the strict requirements on the external PL efficiency and makes observation of the laser cooling effect practically feasible.

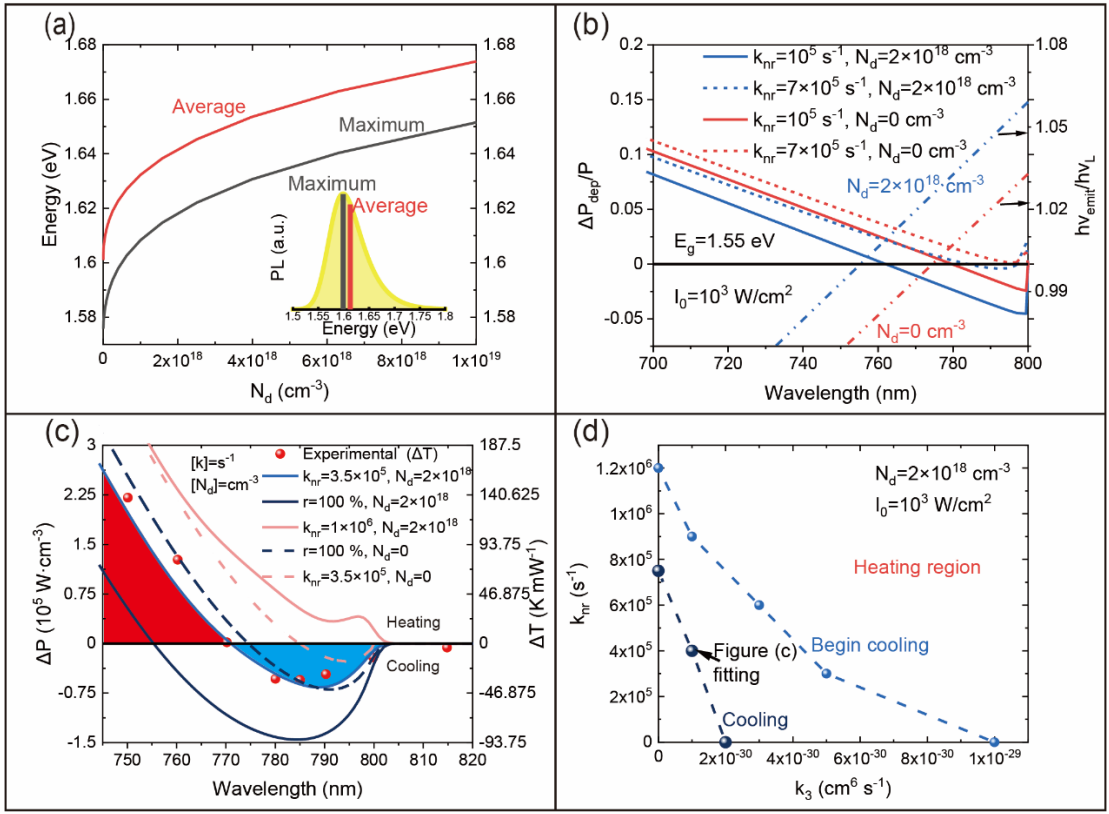

Figure 9. Laser cooling in HOIPs. (a) dependence of the calculated average and maximum PL photon energy vs. concentration of BANDs. (b) spectral dependencies of the cooling efficiency and the ratio of the emitted to absorbed photon energies for two concentrations of BANDs, $N_d = 0$ and $2 \times 10^{18}\ cm^{-3}$, and two non-radiative recombination constants, $k_{nr} = 10^5$ and $7 \times 10^5\ s^{-1}$. (c) Calculated spectral dependencies of the power density

deposited by laser light with the photon power flux of $I_0 = 10^3 \, Wcm^{-2}$ for $N_d = 0$ and $2 \times 10^{18} \, cm^{-3}$ and different values of the non-radiative recombination constant, $k_{nr}$. The experimental laser light induced change of the sample temperature adopted from ref[22] is well explained by the calculated power density for $N_d = 2 \times 10^{18} \, cm^{-3}$, $k_{nr} = 3.5 \times 10^5 \, s^{-1}$ and $k_3 = 10^{-30} \, cm^6 s^{-1}$ corresponding to the external radiative recombination efficiency of 0.98. (d) the relationship between $k_{nr}$ and $k_3$ corresponding to the transition from cooling to heating regime defined by the condition $\Delta P_{dep} \geq 0$ (see Eq. (31)) at all wavelength (blue dashed line). The light blue dashed line represents the relationship for the external PL efficiency of about 0.95. The dark blue dashed line represents the relationship for the external PL efficiency of about 0.98.

As is seen from Eq. (32) the carrier generation rate depends on the power of the illumination photon flux, $I_0$, which we assume to be equal to $1000 \, Wcm^{-2}$. Calculated dependencies of the cooling efficiency given by Eq. (33) along with the ratios of the emitted to absorbed photon energies on the excitation laser photon energy are shown in Figure 9(b). The upper limit for the laser photon energies to produce cooling effect is increasing from 1.59 eV (779 nm) for $N_d = 0$ to 1.63 eV (762 nm) for $N_d = 2 \times 10^{18} \, cm^{-3}$. Even more importantly, increasing the BANDs concentration makes the cooling effect much less sensitive to the non-radiative recombination. Thus, the results in Figure 9(b) show that for $N_d = 2 \times 10^{18} \, cm^{-3}$ a significant cooling is still observed for $k_{nr} = 7 \times 10^5 \, s^{-1}$ whereas this relatively low non-radiative recombination completely eliminates the cooling effect for $N_d = 0$ case.

The photon energy dependence of the laser cooling effect has been calculated using Eq. (31). The results in Figure 9(c) show that the power density extracted or deposited in the sample calculated for the BANDs concentration of $N_d = 2 \times 10^{18}\ cm^{-3}$ and the non-radiative recombination processes with $k_{nr} = 3.5 \times 10^5\ s^{-1}$ and $k_3 = 10^{-30}\ cm^6 s^{-1}$ is in a reasonably good agreement with the spectral dependence of the measured temperature change[22]. It is worth noting that these conditions correspond to an external PL efficiency of about $r = 0.98$ whereas, as is seen in Figure 9(c) and was demonstrated previously[22], almost perfect external PL efficiency of $r \cong 1$ is required to explain the observed cooling effect using a standard semiconductor model[48] with $N_d = 0$.

Although the BANDs-induced anti-Stokes shift improves the efficiency of the laser cooling, the observation of the effect still requires higher than 95% external PL efficiency. This requirement sets limits on the efficiency of the non-radiative recombination processes. Thus, as is shown in Figure 9(d) the cooling sets the upper limits on the non-radiative recombination constants, and requires that $k_{nr} < 1.2 \times 10^6\ s^{-1}$ for the monomolecular and $k_3 < 1 \times 10^{-29}\ cm^6 s^{-1}$ for the Auger non-radiative recombination constant. Note that although this limit of $k_3$ is well below the sometimes quoted values of about $10^{-28}\ cm^6 s^{-1}$ for the Auger recombination constants for HOIPs[27], it is more in agreement with the values of the Auger recombination constants reported for standard semiconductors[50].

6. Conclusions

In this paper we have presented a comprehensive explanation of the origin of extraordinary photophysical properties of hybrid organic-inorganic perovskite materials. It is based on the concept of bistable amphoteric native defects (BANDs) which has been used previously to understand the mechanism of the p-i-n junction formation in undoped perovskite PV absorbers[12]. There are several key component features of the model: (1) the capture and separation of photoexcited electrons (holes) by Coulomb fields of the BANDs in the donor (acceptor) configuration divides the photoexcited carriers into high energy bright and low energy dark fractions with the dark electrons spatially separated from the dark holes and the bright carriers separated from the dark carriers in the k-wavevector space; (2) the spatial separation eliminates radiative and reduces non-radiative recombination of the dark electrons and holes whereas the k-space separation eliminates radiative recombination between the bright and dark carriers. The reduced recombination and the partial separation of the electron and hole transporting regions are essential for explaining exceptional performances of HOIPs based PVs; (3) instantaneous equilibrium of neutral BANDs between donor and acceptor configuration prevents space charge accumulation effects by maintaining local balance between electrons and holes even in structurally and optically inhomogeneous polycrystalline thin films; and (4) the radiative recombination between bright electrons and bright holes produces blue shifted PL that results in a strong laser cooling effect.

All the presented quantitative considerations of the photophysical properties of HOIPs were, for clarity, limited to the prototypical perovskite material, MAPbI$_3$. An obvious question arises to what extent these concepts are applicable to other HOIPs or their alloys. It is an important issue since it has been shown that the best PV performance is achieved through partial substituting of iodine with small amounts of Br and/or Cl and the methylammonium (MA) site with formamidinium (FA)[51]. In both instances the alloying has a direct effect on the strengths of the bonds between the halide and the organic components of the material affecting the formation energy as well as the transformation dynamics between the donor and acceptor configurations of BANDs. Thus, it can be argued that the optimized alloy composition corresponds to a stabilized concentration of BANDs that is most beneficial for performance of the perovskite photovoltaics.

Supporting Information. k-space separation of the bright and dark carriers (Figure S1), dark to bright carrier mobility ratio (Figure S2), time dependent photophysical properties at different time scales (Figure S3, Figure S3, Figure S5, Figure S6), thickness dependence of the average defect concentration in platelets (Figure S7).

Acknowledgements

This work (JWA, WW) was supported by the Singapore National Research Foundation through the Intra-CREATE Collaborative Grant NRF2018-ITC001-001.

B. W. acknowledges the support from the National Natural Science Foundation of China (NFSC) (grant No. 51802331), Science and Technology Program of Guangzhou (No. 2019050001).

T. C. S. acknowledges the support of Nanyang Technological University under its internal grant (M4082480); and the National Research Foundation (NRF) Singapore under its Competitive Research Program (NRF-CRP14-2014-03) and its NRF Investigatorship (NRF-NRFI-2018-04).

# Supplementary Information for

# The Bright Side and the Dark Side of Hybrid Organic-Inorganic Perovskites

*Wladek Walukiewicz[1,2] \*, Shu Wang[3], Xinchun Wu[4], Rundong Li[3], Matthew P. Sherburne[5], Bo Wu[6], Tze Chien Sum[7], Joel W. Ager[1,2,5], and Mark D. Asta[2,5]*

[1]Berkeley Educational Alliance for Research in Singapore (BEARS), Ltd., 1 Create Way, 138602, Singapore

[2]Materials Sciences Division, Lawrence Berkeley National Laboratory, Berkeley, CA, 94720, USA

[3]College of Materials Science and Opto-Electronic Technology, University of Chinese Academy of Sciences, Beijing, 100049, P. R. China

[4]Department of Chemistry Science, University of Chinese Academy of Sciences, Beijing, 100049, P. R. China

[5]Department of Materials Science and Engineering, University of California, Berkeley, California 94720, USA

[6]Guangdong Provincial Key Laboratory of Optical Information Materials and Technology & Institute of Electronic Paper Displays, South China Academy of Advanced Optoelectronics, South China Normal University, Guangzhou 510006, P. R. China

[7]Division of Physics and Applied Physics, School of Physical and Mathematical Sciences, Nanyang Technological University, 21 Nanyang Link, Singapore 637371

*Phone: +1 510 486 5329, Email: w_walukiewicz@lbl.gov

## I. k-space separation of the bright and dark carriers

The average energy, $E_{av}$, of the bright electrons (holes) relative to the conduction (valence) band edge is:

$$E_{av} = \frac{\int_0^\infty (E)\left[V_{br}(E)E^{\frac{1}{2}}f_n(E/k_BT)\right]dE}{\int_0^\infty \left[V_{br}(E)E^{\frac{1}{2}}f_n(E/k_BT)\right]dE} \qquad (S1)$$

The average energy of dark electrons (holes) is given by an analogous expression with $V_{br}(E)$ replaced by $V_{dr}(E)$. The k-wavevectors corresponding to the average energies of the bright or dark carriers are determined form

$$E_{av} = \frac{\hbar^2 k_{av}^2}{2m_n^*} \qquad (S2)$$

**Figure S1** shows the dependencies of the $|k_{av}|$ as well as the difference of the $|k_{av}|$ on the concentration of BANDs for the bright and dark carriers. The calculations were performed for the nondegenerate carrier statistics i.e. for the total free carrier concentration of less than $2 \times 10^{18}\ cm^{-3}$. It is seen that the difference between $|k_{av}|$ for the bright and dark carriers is larger than $3 \times 10^8\ m^{-1}$ and is only very weakly dependent on the concentration of BANDs.

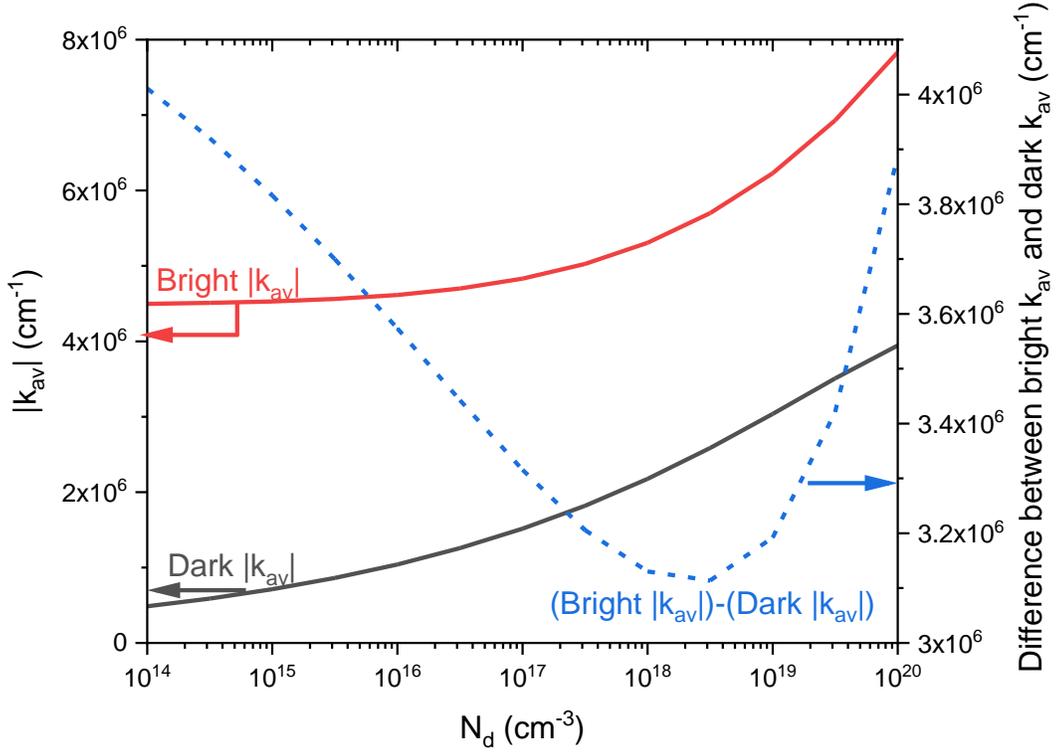

**Figure S1** Average k-wavevector value $|k_{av}|$ for bright and dark carriers as well as the difference of the $|k_{av}|$, as a function of the concentration of BANDs. Assumes Boltzmann statistics and parabolic dependence of energy and k-wavevector.

## II. Dark to bright carrier mobility ratio

As is discussed in section 4.1 of the main text the photoconductivity (PC) of photoexcited electrons and holes is proportional to $2(n_{br}\mu_{br} + n_{dr}\mu_{dr})$. Therefore, it depends on the dark to bright carriers' mobility ratio, $\mu_{dr}/\mu_{br}$. To estimate the mobility ratio we assume that charge scattering is dominated by two mechanisms, optical phonon and ionized

BANDs scattering. The relaxation time approximation has been used for the optical phonon scattering, with the microscopic relaxation time given by[1],

$$\frac{1}{\tau_{op}(E)} = \frac{e^2 \omega_{LO}}{4\sqrt{2}\pi\epsilon_0 \hbar}\left(\frac{1}{\epsilon_\infty} - \frac{1}{\epsilon_0}\right)\frac{\sqrt{m^*}}{\sqrt{E}} \times [(n_q + 1)[\sqrt{1 - \frac{\hbar\omega_{LO}}{E}} + \frac{\hbar\omega_{LO}}{E}\sinh^{-1}\left(\frac{E}{\hbar\omega_{LO}} - 1\right)^{\frac{1}{2}}] + $$
$$n_q[\sqrt{1 + \frac{\hbar\omega_{LO}}{E}} - \frac{\hbar\omega_{LO}}{E}\sinh^{-1}\left(\frac{E}{\hbar\omega_{LO}}\right)^{\frac{1}{2}}] \quad (S3)$$

where E is the electron energy, $\omega_{LO}$ is longitudinal optical phonon frequency, $\epsilon_\infty$ and $\epsilon_0$ are high frequency and static dielectric constants, and $n_q$ is the Bose-Einstein LO phonons distribution function.

The charged defect scattering is described by the microscopic relaxion time of the form (see e.g. reference[2]),

$$\frac{1}{\tau_{def}(E)} = 2.415 \times \frac{N_{def}}{\epsilon_0^2}\left(\frac{m^*}{m_0}\right)^{-1/2}(xT)^{-3/2} \times [\ln(1 + \frac{4x}{a}) - \frac{4x/a}{1+4x/a}] \quad (S4)$$

where $x = E/k_B T$, $N_{def}$ is the total concentration of the charged defects and $a$ is the reduced screening energy, which is given by

$$a = \frac{\hbar^2}{2ml_D^2 k_0 T}$$

$$\frac{1}{l_D^2} = 5.80 \times 10^{13} \times \frac{\left(\frac{m^*}{m_0}\right)^{\frac{3}{2}} T^{\frac{1}{2}}}{\epsilon_0} F_{-\frac{1}{2}}(\eta)$$

and $F_{-\frac{1}{2}}(\eta)$ is the Fermi-Dirac integral of order $-\frac{1}{2}$. To calculate the total mobilities of the dark and bright carriers the combined relaxation time $\tau^{-1}(E) = \tau_{op}^{-1}(E) + \tau_{def}^{-1}(E)$ has to be averaged over the distribution for both types of carriers.

Thus, for bright carriers,

$$\mu_{br} = \frac{e}{m^*}\int_0^\infty \tau_{br}(E) V_{pl}(E) E^{\frac{1}{2}} f_n(z) / \int_0^\infty V_{pl}(E) E^{\frac{1}{2}} f_n(z) \quad (S5)$$

With an analogous formula for $\tau_{dr}$ but with $V_{pl}(E)$ replaced by $1 - V_{pl}(E)$.

**Figure S2** shows the calculated dependence of the mobility ratio on the total carrier concentration for three different concentration of BANDs. The calculations were performed using the values $\varepsilon_0 = 18$ [3], $\varepsilon_\infty = 7$ [4] and $\hbar\omega_{LO} = 11.5\ meV$ [5]. The ratio is very weakly dependent on the concentration of BANDs and is relatively constant for the nondegenerate carrier statistics i.e. for the charge carrier concentrations smaller than $2 \times 10^{18}\ cm^{-3}$. In our PC calculations we have adopted the value $\frac{\mu_{dr}}{\mu_{br}} = 0.14$ that falls in the range of the calculated values for the BANDs and charge carrier concentrations expected for typical photoconductivity experiment in thin polycrystalline films.

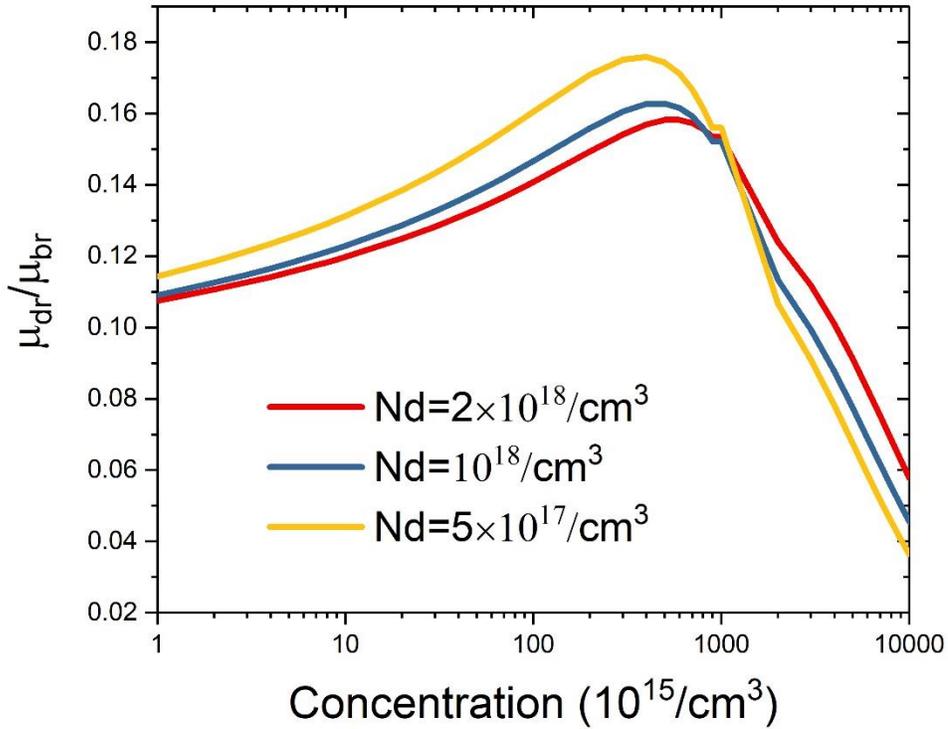

**Figure S2** Calculated dependence of the dark to bright carrier mobility ratio on the total carrier concentration for three different concentrations of BANDs. The concentration of the ionized scattering centers $N_{def} = N_D^+ = N_A^- = 2 \times N_d$.

### III. Time dependent photophysical properties at different time scales

As is shown in the section 4 of the main text the separation of photoexcited carriers into bright and dark fractions has a large effect on the optical properties of HOIPs. The calculations presented in the main text are focused on the transients of the photoeffects in the time range from 0 to 300 ns. In this section we examine the transients for times shorter than 10 sec and for much longer times up to 3000 ns. These calculations aim at demonstrating how the capture of the photoexcited carriers by BANDs affects the transients of these photoeffects at so much different time scales. For consistency with main text calculations we have adopted the values of $k_2 = 8 \times 10^{-11}\ cm^3 s^{-1}$ for the bimolecular radiative recombination, $k_{cp} = 2 \times 10^{-11}\ cm^3 s^{-1}$ for the carrier capture rate by BANDs and $k_{nr} = 10^6\ s^{-1}$ for the monomolecular nonradiative recombination rate. **Figure S3** presents calculated transients of the total and dark carrier concentrations (a), photoluminescence (b) and photoconductivity (c) for times shorter than 10 ns. It is seen that adding BANDs increases the dark carrier concentration resulting in a slower decay of $n_{tot}$ or (PB) and significantly faster decay of PL and PC.

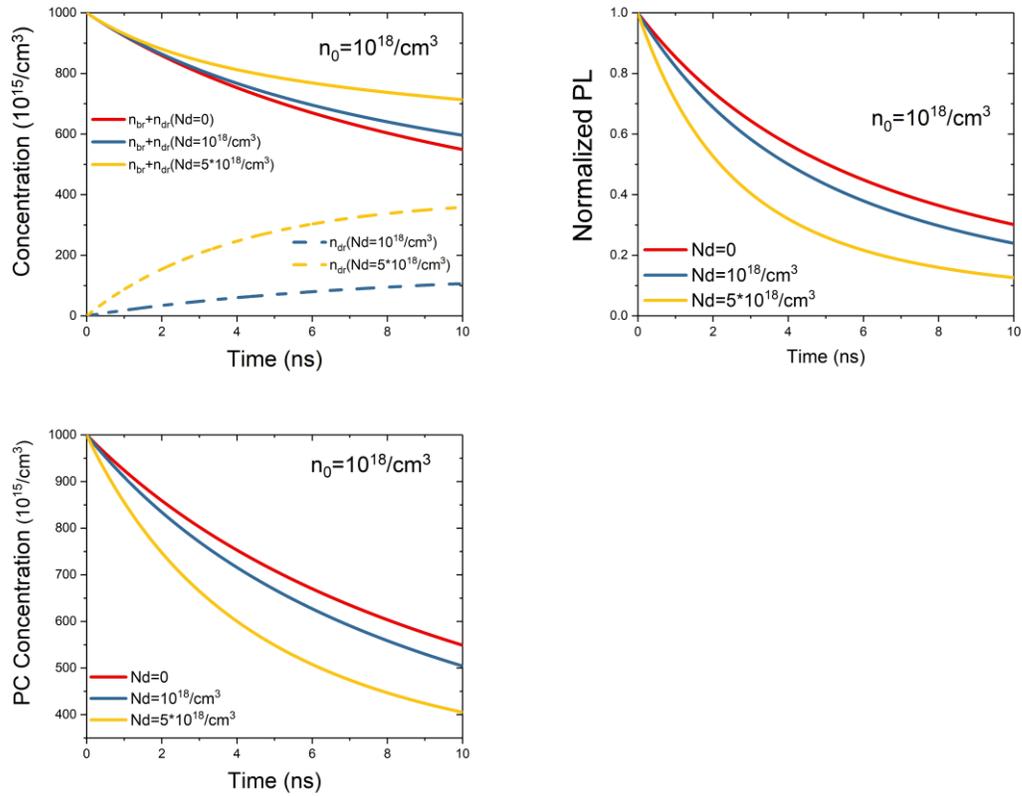

**Figure S3** Initial transients of $n_{tot}$ (PB) and the dark carrier concentration (a), PL (b) and PC (c) for the photoexcitation density $n_0 = 10^{18}\ cm^{-3}$ and different concentrations of BANDs. The capture of the carriers by BANDs results in an increase in the dark carrier concentration (a) leading to slower decay of (PB) and faster decays of PL (b) and PC (c). The protective role of BANDs is already visible at these short times where, as is seen in (a) the slowest temporal decay of the total concentration is found for the highest concentration of BANDs.

**Figure S4** shows the transients calculated for times up to 3000 ns with the same set of parameters as that used in **Figure 2** in the main text. The protective role of BANDs is clearly visible in all three photoeffects which show larger values for larger concentrations of BANDs.

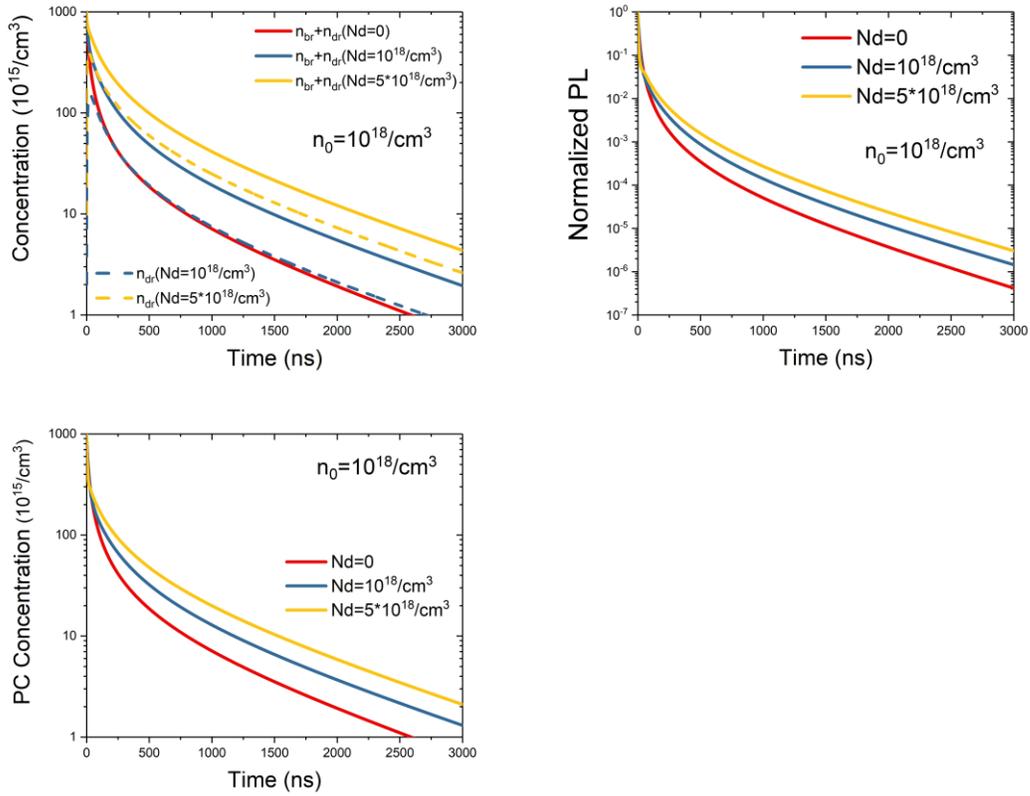

**Figure S4** Calculated long time transients for the photoexcitation density $n_0 = 10^{18}\ cm^{-3}$ and different concentration of BANDs. (a) the total and dark carrier concentration (a); PL (b) and PC (c). The dark to the total concentration ratio attains a constant, equilibrium value at these long times.

**Figures S5** and **S6** present the short (< 10 ns) and long (< 3000 ns) transients for $N_d = 1 \times 10^{18}\ cm^{-3}$ and three different photoexcitation densities, $n_0$. The initial decays (**Figure S5**) clearly show that the carrier capture by BANDs is responsible for the fast decay of the PC and an even faster decay of the PL. The decay rates increase with increasing excitation density, $n_0$.

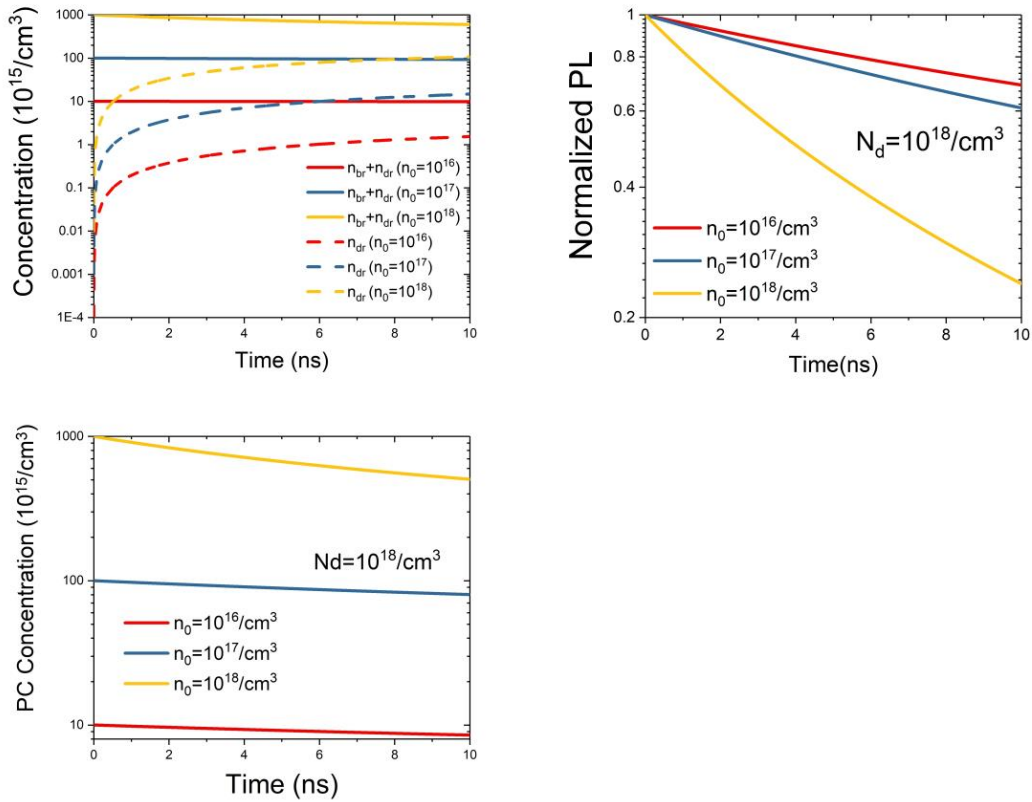

**Figures S5** Calculated transients of the total and dark carrier concentration (a) PL (b) and PC (c) for the concentration of BANDs $N_d = 1 \times 10^{18}\ cm^{-3}$ and three different initial photoexcitation carrier densities. The dark to the total carrier concentration ratios reaches equilibrium at about 10 ns.

**Figure S6** presents the calculated long-time transients for $N_d = 1 \times 10^{18}\ cm^{-3}$ and three different photoexcitation densities. It is seen that the initial decays are strongly dependent on the photoexcitation level. This is clearly evident in the PL transients in **Figure S6**(b) where for the highest photoexcitation level, $n_0 = 10^{18}\ cm^{-3}$ the PL intensity drops by few orders of magnitude in less than 500 ns whereas only a modest, less than order of magnitude decay is seen for $n_0 = 10^{16}\ cm^{-3}$. These results indicate that the rate of the initial rapid PL decay is not a reliable measure of the importance of nonradiative recombination and that it is extremely important to know the photoexcitation level for any quantitative interpretation of experimental time resolved photoeffects.

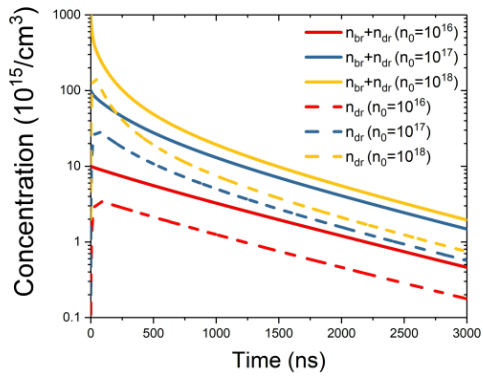
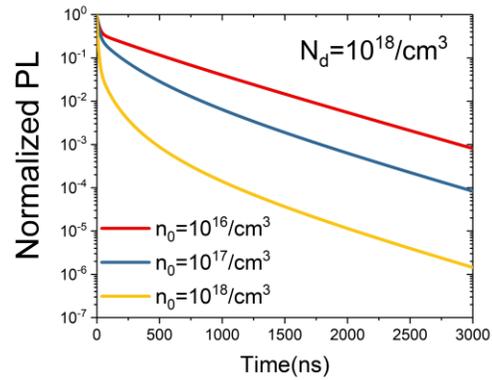
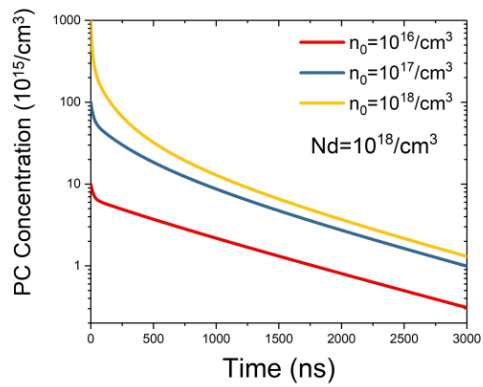

**Figure S6** Calculated time resolved total carrier concentration (PB) and the dark carrier concentration (a), PL (b) and PC (c) for the BANDs concentration $N_d = 10^{18}\ cm^{-3}$ and different values of the initial photoexcitation density, $n_0$. The results show that the decay rates of all the photoeffects depend on the excitation intensity $n_0$.

### IV. Thickness dependence of the average defect concentration in platelets

In the **Figure 8** (c) and (d) of the main text, we show how the thickness dependent PL peak energy and intensity measured in MAPbI$_3$ single crystal platelets can be explained by different concentrations of BANDs[6]. Here we demonstrate that the dependence of the concentration of BANDs on the platelet thickness, $d$ can be understood assuming low concentration of BANDs in the bulk and much higher concentration in a thin subsurface layer. Under these assumptions the average concentration of BANDs is given by the expression:

$$N_d(d) = N_{db} + \frac{(N_{ds} - N_{db}) \times d_s}{d} \qquad (S6)$$

where $N_{db}$ is the bulk BAND concentration and $N_{ds}$ is the concentration of BANDs in the surface layer of thickness $d_s$. As is shown in **Figure S7** the best fit for the average $N_d(d)$ given by Eq. (S6) is obtained with $N_{db} = 1 \times 10^{15}\ cm^{-3}$ and $N_{ds} \times d_s = 6.2 \times 10^{13}\ cm^{-2}$. Such a good fit with a single set of parameters indicates a reproducible growth conditions with both bulk and surface concentration of BANDs independent of the platelet thickness.

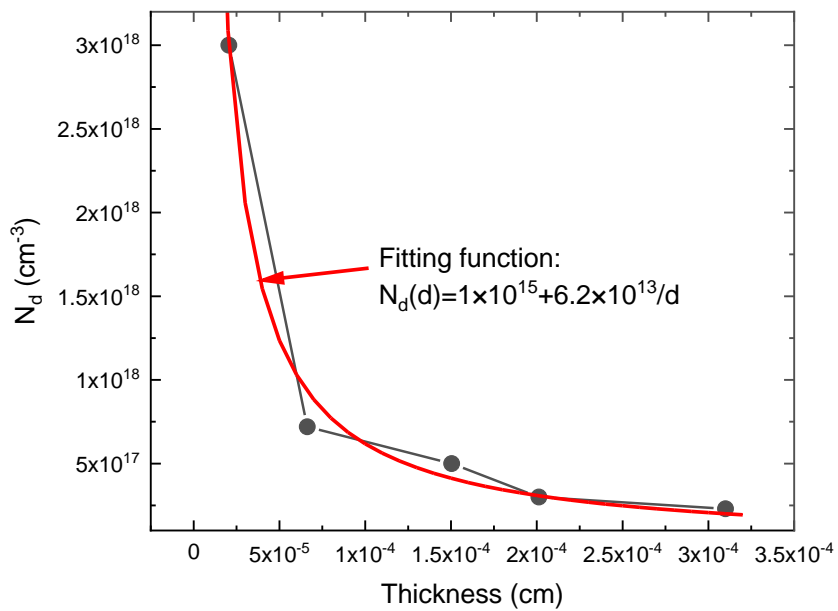

**Figure S7** Average concentration of BANDs vs. platelet thickness. The points represent the concentrations determined from the PL spectra and listed in **Figure 8** (c) (main text)[6]. The fit (red line) is the concentration averaged over the bulk and the surface layer using Eq. S6.